\documentstyle[preprint,aps,psfig]{revtex}
\def\Journal#1#2#3#4{{#1} {\bf #2}, #3 (#4)}

\def\AP{{Ann. Phys.}}
\def\APF{{Ann. Phys.} Fr.}
\def\CPC{{Comp. Phys. Com.}}

\def\PRL{Phys. Rev. Lett.}
\def\PREV{Phys. Rev.}
\def\PRD{{Phys. Rev.} D}
\def\PR{{Phys. Rep.}}
\def\APJ{{Astrophys. J.}}
\def\APJS{{Astrophys. J. Supp.}}
\def\AA{{Astron. Astrophys.}}
\def\NAT{{Nature}}
\def\NPA{{Nucl. Phys.} A}

\def\RMF{{Rev. Mex. F\'\i s.}}
\def\JCP{{Comput. Phys}}

\def\be{\begin{equation}}
\def\ee{\end{equation}}
\def\bea{\begin{eqnarray}}
\def\eea{\end{eqnarray}}
\def\bm{\boldmath}

\begin{document}

\title{Dynamics of spherically symmetric spacetimes: Hydrodynamics and Radiation}

\author{ Marcelo Salgado \footnote{e-mail address: marcelo@nuclecu.unam.mx}\\ \small {\it Instituto de Ciencias Nucleares }
\\ \small {\it Universidad Nacional Aut\'onoma de M\'exico} \\ 
\small {\it Apdo. Postal 70--543 M\'exico 04510 D.F., M\'exico}}
\maketitle

\abstract{Using the 3+1 formalism of general relativity we obtain the 
equations governing the dynamics of spherically symmetric spacetimes 
with arbitrary sources. We then specialize for the case of perfect fluids 
accompanied by a flow of interacting massless or massive particles 
(e.g. neutrinos) which are described in terms 
of relativistic transport theory. We focus in three types of coordinates: 
1) isotropic gauge and maximal slicing, 2) radial gauge and polar slicing, and 
3) isotropic gauge and polar slicing.}

\vskip 2cm
Submitted to Phys. Rev. D 

\newpage

\section{Introduction}
One of the most fascinating phenomena in gravitational physics is that 
of gravitational collapse. Notably, the gravitational collapse of astrophysical 
bodies culminating in the formation of black holes. 

The historical controversy on the final fate of gravitational collapse of 
compact objects such as white dwarfs and neutron stars raised by the 
discovery of maximum mass limits and the subsequent 
 stability analysis led to find, at least for the most 
ideal configurations, definite answers and concrete predictions 
depending on the initial conditions and the equation of state 
of matter (see Refs. \cite{HWW,ST83} for a review). The simplest situation 
describing the gravitational collapse ending in a black hole formation 
is that of a spherical ball of 
pressureless and homogeneous fluid (the well known Oppenheimer-Snyder 
dust collapse \cite{OS}). 
The solution shows the appearance of an {\it event horizon} revealing 
thus the formation of a black hole after a finite proper time. 

Since that pioneering investigation, much has been done for more complicate 
and realistic initial configurations. On one hand, the accelerated 
development in the area of computation and the recent advances on the numerical 
analysis of Einstein equations  have made possible computing 
the dynamics of rather complex spacetimes faster and for longer term evolutions. 
On the other hand, the advances in particle and nuclear physics have led 
to a better knowledge of the conditions of matter at high densities, providing then 
more realistic models for the matter in cores and neutron stars.

Most of the recent analysis of gravitational collapse leading to a 
black hole formation have been done in spherical symmetry 
and following a more ``modern'' point view in light of the 
3+1 formulation of general relativity and other formulations better adapted to numerical 
stability (see Ref.\cite{Alcubierre,Kelly} and the references therein). 
Among these investigations, 
there is the one of Shapiro \& Teukolsky \cite{ST80} who studied the 
collapse of polytropes by imposing an isotropic gauge and maximal 
slicing coordinates, with initial 
conditions provided by the Tolman-Oppenheimer-Volkoff equation of 
hydrostatic equilibrium. Later, a similar study, was performed by 
Schinder, Bludman and Piran \cite{SBP} in comoving coordinates with a
polar slicing, and by Gourgoulhon \cite{gour1,gour2} in radial gauge and 
polar slicing, this latter improving previous analysis by the incorporation 
of realistic equations of state for the nuclear matter.

The effects of the coordinate choice and the slicing condition on the 
time evolution has been exhibited for two of the most popular choices 
(isotropic and radial gauges with maximal and polar slicings) 
by comparing with the analytical solution of Oppenheimer-Snyder 
(OS) \cite{gour2,STP85,STP86}. For the time being, the discussion 
of gauges is postponed to Sec. VII.
\bigskip

The analysis of gravitational collapse of compact stars and iron cores would not 
be complete if the influence of neutrinos were not taken into account. This has 
been recognized by a long list of authors since the precursor investigations 
of Colgate and White \cite{CW}, and May and White \cite{MW}, 
who analyzed the effect of neutrinos in supernovae explosions. 
Later Wilson \cite{wilson}, performed full 
general relativity computations with a neutrino flow described in 
terms of the relativistic Boltzmann equation. Wilson's analysis included 
electron and muon massless neutrinos assuming that the 
corresponding antineutrinos contributed in the 
same basis. The interaction of neutrinos with the star's fluid was 
described by an opacity function. Unlike previous studies, Wilson's 
found that the heat conduction by neutrinos is not sufficient to 
eject any material from a collapsing star. 
In all those studies black hole formation were not analyzed 
but only evolution configurations terminating in stable states corresponding 
to white dwarfs or neutron stars. 

An updated analysis 
were carried out by Mayle, Wilson and Schramm \cite{MWS} using a Boltzmann 
code and for a large set of mass configurations. Neutrino signals from various species 
were analyzed within time scales of $\sim 1$ s after the supernova explosion. 
It is to be mentioned that previously Saenz and Shapiro \cite{SS} 
had computed a non-spherical Quasi-Newtonian collapse acompanied by neutrino and 
gravitational radiation.

Burrows and coworkers have also analyzed during the last twenty years the 
mechanism of Type II supernovae (SNII) explosions and the role of neutrinos 
(see \cite{BHF,Burrows88} and references therein). Among these investigations, 
we find an interesting model of long term neutrino emission from the hot 
protoneutron star phase to the final outcome of a stable cold neutron star 
\cite{BL}.

Many other recent investigations have confirmed and improved in several aspects previous 
findings on SNII (see \cite{MM,Mezzacapa,BBM,Lieben1,Lieben2,MB} and 
references therein). 
For the case of a core collapse leading to a black hole two scenarios are recognized 
\cite{BBM}. 
One called {\it early} black hole formation which 
is generically associated with accreting protoneutron stars which form
from the collapse of degenerate cores of massive stars 
\cite{MB,Burrows}. The 
accretion of some tenths to one solar mass can last a second, and the 
exceeding of the maximum mass drives the protoneutron star into a black 
hole collapse in a typical time scale of $\sim 0.5$ ms. In this 
scenario the neutrino signal is abruptly 
cut off after the black hole forms, and the typical neutrino luminosities prior the 
cutoff are $\sim 10^{52}$ erg/s per flavor.

The second scenario called {\it late} 
 black hole formation typically arises by a softening of the high-density 
equation of state of the protoneutron star 
\cite{Baum0,Baum1,Janka,Baum2}. The phase transition from 
the neutron star matter to a more exotic state which include kaon condensates 
\cite{TPL,BB}, or hyperon condensates \cite{Glend} can lower the maximum 
allowable mass to $\sim 1.5 M_\odot$ 
\cite{Baum1,Baum2}, driving thus a stable protoneutron star to an 
unstable regime and finally to a collapse into a black hole. This kind of core collapse 
can last $\sim 10$ s before the cutoff and the luminosity of neutrinos 
is ten times lower than the luminosity of the early case.

It is encouraging to note that for a SNII at a distance of $\sim 10$ kpc 
which explodes within the early scenario, SuperKamiokande can probe $\bar \nu_e$ 
masses down to 1.8 eV by comparing the arrival times of high and low energy 
neutrinos within the reaction $\bar \nu_e + p\rightarrow e^+ +  n$ in a 
Cerenkov detector (see Ref. \cite{BBM} for details).

In fact the very recent announcement on the measurements of solar 
neutrinos from the decay of $^8{\rm B}$ by the Sudbury Neutrino Observatory 
(SNO) \cite{SNO} via charged current 
interactions and by the elastic scattering of electrons reveals 
that neutrinos could be changing flavor as they travel from the sources to the 
Earth. This discovery if confirmed could corroborate the oscillating behavior of 
neutrinos and therefore their massive nature. The fluxes measured of the 
different flavors are in close agreement with the predictions of the solar models. 
The SNO experiment then implies 
that the upper limit of the mass squared difference between the 
$\nu_e$ and the $\nu_\tau$ or  $\nu_\mu$ is less than $10^{-3}\,{\rm eV}^2$ \cite{SNO}. 
This result when combined with the current bounds on $m_{\nu_e}$ of $2.8$ eV 
and $\Delta m_{\nu_\mu\nu_\tau}^2$ (assuming neutrino oscillations) 
provides a limit for the sum of the masses 
of the three neutrino species in the range $[0.05,8.4]$ eV \cite{SNO}.

One proposal to measure the $\nu_\tau$ and  $\nu_\mu$ masses indirectly 
and that can corroborate the SNO findings is the one which uses a 
time-of-flight technique \cite{BBM}, for neutrinos emitted in the early black 
hole formation scenario discussed above. The point is that if neutrinos 
are massive then there is a delay (relative to a massless neutrino) in the 
cutoff of the neutrino signal as measured on Earth after the black hole 
forms, and it is given by $\Delta t\sim (m_\nu/E_\nu)^2$ for distances 
of $\sim 10$ kpc. This delay can affect the event rate measured 
in a detector. The conclusion is that assuming luminosities 
$L\sim 10^{52}$ erg/s per flavor at the cutoff time, 
SuperKamiokande can probe e-neutrino 
masses as small as 1.8 eV for $T_{\nu_e}\sim 3.5$ MeV, whereas 
the OMNIS \cite{OMNIS} or SNO detector can detect $m_{\nu_\mu,\nu_\tau}$ masses as 
small as 6 eV for $T_{\nu_\mu,\nu_\tau} \sim 8$ MeV.
\bigskip

A collapse scenario which is rather different from the above consists in the 
accretion of matter by an old neutron star near the maximum mass limit. 
Gourgoulhon and Hansel \cite{gour3} have analyzed the neutrino emission 
during the collapse to a black hole within this scenario 
via nonequilibrium $\beta$ processes, assuming that the nuclear matter 
is transparent for neutrinos (i.e., opacities were neglected). 
Instead of using neutrino transport, a 
{\it regularized} geometrical-optics model adapted to massless neutrinos 
was adopted. This model was thus intended to provide upper bounds 
in the neutrino burst. The collapse lasts typically 
a millisecond (the time that takes place 
the black hole formation), and in the most favorable conditions the total 
energy of $\bar \nu_e$ and $\bar \nu_\mu$ antineutrinos is $\sim 10^{51}$ erg, while the 
energy of the corresponding neutrinos is several orders of magnitude lower. 
This is even lower for the $\nu_\tau$ and $\bar \nu_\tau$ 
neutrinos. The main conclusion 
is that a collapse of this kind at a distance of $\sim 10$ kpc would be 
undetectable by the current neutrino detectors.
\bigskip

Finally, another scenario which has been analyzed in the past 
is the dynamics of collisionless gas of particles which mimic spherical star clusters, and the 
possibility of a cluster collapse into a supermassive black holes 
\cite{ST85b,ST85c,ST86,STK,RST}. 
The motivation was to provide a theoretical description for the 
formation of supermassive black holes that could exist in the centers of 
galaxies. Recently, a similar study which include gamma-ray bursts, 
is the one analyzed by 
Linke {\it et al.} \cite{Linke}, where the collapse of supermassive stars 
($M\sim 10^5 M_\odot - 10^9 M_\odot$) with emission of thermal neutrinos 
is considered. In that work, the spacetime is foliated by outgoing null hypersurfaces 
rather than using a 3+1 foliation of spacetime. 
\bigskip

In view of the different scenarios of gravitational collapse available today 
and the miscellaneous predictions within each of them it is worth  
pursuing the investigations along these lines. Only in this way there will 
be at hand a large set of models which the forthcoming (e.g. \cite{OMNIS}) and 
recent observations \cite{SNO} will validate or rule out.

Although the paper is written in the same spirit of various papers which deal with 
the system of equations Einstein-Hydrodynamics-Boltzmann, several 
aspects distinguish from them. For instance, most formalisms treat neutrinos as massless 
particles, except perhaps the one of Harleston and collaborators \cite{HV,HH}. 
Here massive particles are considered from the onset and previous equations are 
recovered in the massless limit. The relativistic Boltzmann equation is written in 
terms of 3+1 variables for generic spacetimes. This had done in the past only in 
spherical symmetry. Therefore, the relativistic Boltzmann equation presented 
here is coupled from the 
onset to the 3+1 Einstein's equations. That is, the curvature effects appear in terms of 
the lapse, the shift, the extrinsic curvature and the three-metric. 
The hydrodynamic equations are derived also in the context of the 3+1 formalism 
and they couple to the neutrinos via collision integrals. In particular the equation for 
the velocity field of the fluid is written in several forms each of one 
is useful whether one uses different numerical methods. At this 
regard, a general relativistic Euler equation is presented using the 
tetrad formalism. Its quasi-Newtonian 
form allows an easy interpretation of several terms, and reduces to well known 
equations in various limits. Such an Euler equation turns to be better 
adapted for spectral methods than the equation for the momentum current density 
\cite{gour1,gour2}. The system of equations are then specialized for spherical 
symmetry and written in three different coordinates: 
1) isotropic gauge and maximal slicing, 2) radial gauge and polar slicing, and 
3) isotropic gauge and polar slicing.

This is the first of a series of papers where the gravitational 
collapse of various kinds of matter will be analyzed. 

The paper is organized as follows: in Sec. II we present succinctly 
the 3+1 formalism of general relativity rather more to fix the notation than 
to give a detailed description. In Sec. III we consider the case of two 
interacting sources of matter: a perfect 
fluid and a flow of relativistic particles described in terms of 
relativistic transport theory. Section IV treats the relativistic 
transport theory. Section V deals with therodynamics.
In section VI spherical symmetry is considered and in 
section VII three coordinate choices and 
slicing conditions are analyzed and discussed in light of the previous studies. 
Finally we conclude with some remarks and the plans for the forthcoming investigations 
along this line.

\section{The 3+1 formalism of general relativity}
One of the most popular reformulations of general relativity when 
tackling numerical problems is the 3+1 or Adison-Deser-Misner (ADM) 
formulation. We shall not enter into the details of the derivation of 
the equations (see Refs. \cite{ADM,Choquet,msnotes,york79,Wald}) but rather discuss the general 
idea in order to fix the notations.

The main idea is as follows: under general assumptions (see \cite{Choquet,Wald} and 
reference therein for details) a globally hyperbolic spacetime $(M^4, g_{\mu\nu} )$ 
can be foliated by a family of space-like hypersurfaces 
$\Sigma_t$ (Cauchy surfaces). Each hypersurface represents a Riemannian 
sub-manifold $(M^3, h_{ij})$ endowed by an {\it induced metric} $h_{ij}$ 
(the 3-metric). It is then assumed a local coordinate system $(t,x^i)$ for 
the spacetime, the spatial part $(x^i)$ represents a local coordinate system 
for $\Sigma_t$, while $t$ is a global time function that parametrizes $\Sigma_t$. 
The embedding of $\Sigma_t$ in spacetime is completed 
by the {\it extrinsic curvature} of $\Sigma_t$. This is defined by
\be \label{K}
K_{\mu\nu}:= -\frac{1}{2} {\cal L}_{\mbox{\boldmath{$n$}}}h_{\mu\nu}\,\,\,\,,
\ee
where ${\cal L}_{\mbox{\boldmath{$n$}}}$ stands for the Lie derivative 
along the normal $n^\mu$ to $\Sigma_t$ and $h_{\mu\nu}:= g_{\mu\nu} + 
n_\mu n_\nu$. The vector field $n^\mu$ is 
time-like ($n^\mu n_\mu=-1$) 
and the convention used for its components 
with respect to the coordinate base adapted to the spacetime foliation is 
as follows:
\be
n^\mu= \left(N,N^i\right)\,\,\,.
\ee
This convention means that $n^\mu$ points towards the {\it future}. 
Since  $n^\mu$ is a unit time-like vector, it is customary to interpret 
$n^\mu$ as the four-velocity of the so-called Eulerian observer ${\cal O}_{\rm E}$. 
The {\it scalar} quantity $N$ (the lapse function) represents thus  
the rate at which ${\cal O}_{\rm E}$ sees the flow of its proper-time 
as compared with the intervals between two neighboring hypersurfaces 
$\Sigma_t$ and $\Sigma_{t+dt}$. The 3-vector $N^i$  
(the {\it shift-vector}), represents the coordinate 3-velocity 
at which the Eulerian observer moves with respect to the coordinates 
$(t,x^i)$. In this way, the four-metric reads 
\be\label{3+1metric}
ds^2= -(N^2 - N^iN_i)dt^2 - 2N_i dt dx^i + h_{ij} dx^i dx^j\,\,\,.
\ee
We emphasize that in many studies the convention on the 
sign of the shift vector is different from the one adopted here. 
This has to be taken into account particularly for purposes of 
comparison between the final form of equations in spherical symmetry 
presented here and the corresponding equations of the references which 
employs the opposite sign.

A useful formula for $K_{ij}$ obtained from Eq. (\ref{K}) is 
\be \label{K_ij}
K_{ij} = -\nabla_i n_j= -N\Gamma^t_{ij} = -\frac{1}{2N}\left( 
\frac{\partial h_{ij}}{\partial t} + \,^3\nabla_j N_i + 
 \,^3\nabla_i N_j \right) 
\,\,\,\,.
\ee
where $\,^3\nabla_j$ stands for the covariant derivative compatible 
with $h_{ij}$. This is to be regarded as an evolution equation for the 
three-metric $h_{ij}$.

The trace of the extrinsic curvature is simply given by,
\be \label{divK2}
K:=  -\nabla_\alpha n^\alpha \,\,\,\,.
\ee

Another useful quantity is the acceleration of ${\cal O}_{\rm E}$ given by 
\be\label{eulerac}
a^\mu := n^\nu \nabla_\nu n^\mu = \,^3\nabla^\mu [{\rm ln}N]\,\,\,,
\ee
which allows for the lapse the interpretation of the acceleration potential 
for ${\cal O}_{\rm E}$ \cite{york79}.

The orthogonal decomposition of the energy-momentum tensor
 in components tangent and 
orthogonal to $\Sigma_t$ leads to \cite{york79}:

\be\label{Tort}
T^{\mu \nu} = S^{\mu \nu} + J^{\mu} n^{\nu} + n^\mu J^{\nu} 
+ E n^\mu n^\nu\,\,\,\,.
\ee

The tensor $S^{\mu \nu}$ is symmetric and often called the   
{\it tensor of constraints}; $J^\mu$ is the    
{\it momentum density vector} and    
$E$ is the {\it total energy density} measured by the Eulerian observer
 ${\cal O}_{\rm E}$. Both $S^{\mu\nu}$ and $J^\mu$ are orthogonal to 
$n^\mu$.

We emphasize that for the specific applications we will study, 
$T^{\mu\nu}$ will be the total energy-momentum tensor of matter 
which can be composed by the contribution of different types of sources:
\be\label{Ttot}
T^{\mu\nu}= \sum_i T^{\mu \nu}_i\,\,\,\,.
\ee
This means that
\be\label{3+1mattvartot}
E = \sum_i E_i \,\,\,,\,\,\,
J^\mu = \sum_i J^\mu _i \,\,\,,\,\,\,
S^{\mu\nu} =  \sum_i  S^{\mu\nu}_i \,\,\,\,.
\ee

The projection of Einstein equations $R_{\mu\nu}= 4\pi G_0\left(2T_{\mu\nu}-
T^{\alpha}_{\,\,\,\alpha} g_{\mu\nu}\right)$ 
in the directions  tangent and 
orthogonal to $\Sigma_t$, followed by 
the use of the Gauss-Codazzi-Mainardi equations leads to the 
3+1 form of Einstein equations:

\be\label{CEHf}
^3 R + K^2 -  K_{ij}  K^{ij}= 16\pi G_0 E \,\,\,,
\ee
known as the Hamiltonian constraint.

\be\label{CEMf}
^3 \nabla_l K_{\,\,\,\,\,i}^{l} - \,^3\nabla_i K= 8\pi G_0 J_i 
\,\,\,\,,
\ee
known as the {\it momentum constraint} equations.

Finally, the {\it dynamic} Einstein equations read
\bea
& & 
\partial_t K_{\,\,\,j}^i + N^l \partial_l K_{\,\,\,j}^i +
K_{\,\,\,l}^i \partial_j N^l
- K_{\,\,\,j}^l \partial_l N^i
+ \,^3\nabla^i\,^3\nabla_j N
- \,^3 R_{\,\,\,j}^i N - N K K_{\,\,\,j}^i \nonumber \\
\label{EDEf}
& &
= 4 \pi G_0 N\left[ (S-E)\delta^i_{\,\,j} - 2S^i_{\,\,j}\right]\,\,\,\,
\eea
where $S= S^l_{\,\,l}$ is the trace of the tensor of constraints, and all 
the quantities written with a `$3$' index refer to those computed with 
the three-metric $h_{ij}$. Moreover, 
under the 3+1 formalism tensor quantities tangent to 
$\Sigma_t$ use the three-metric to raise and lower their spatial indices. 
The equations (\ref{K_ij}) and (\ref{EDEf}) are the set of the 
Cauchy-initial-data evolution equations 
for the gravitational field subject to the constraints Eqs. (\ref{CEHf}) and 
(\ref{CEMf}).

An evolution equation for the trace $K$ is obtained by taking the 
trace in Eq. (\ref{EDEf}):
\be\label{EDK}
\partial_t K  + N^l \partial_l K  + \,^3\Delta N
-N\left( \,^3 R  +  K^2\right) 
= 4 \pi G_0 N\left[S-3 E\right] \,\,\,.
\ee
where $\,^3\Delta$ stands for the Laplacian operator compatible 
with $h_{ij}$.
 
This can be simplified by using Eq.(\ref{CEHf}) to give
\be\label{EDK2}
\partial_t K  + N^l \partial_l K  + \,^3\Delta N
-N  K_{ij}  K^{ij} 
= 4 \pi G_0 N\left[S + E\right] \,\,\,.
\ee
 
To complete the system of equations, we consider also the matter equations
\be
\nabla_\mu T^{\mu\nu}=0\,\,\,,
\ee
which according with Eq. (\ref{Ttot}), these can be written as

\be\label{divT}
\nabla_\mu T^{\mu\nu}_\psi = -{\cal F}^\nu\,\,\,,
\ee
where $T^{\mu\nu}_\psi$ is the energy-momentum tensor of 
certain fields that are collectively labeled by $\psi$ and 
${\cal F}_\nu:= \nabla_\mu T^{\mu\nu}_{\rm ext}$ are the ``forces'' 
exerted by the external fields (fields other than $\psi$). 
For instance, we shall 
consider the case where the total energy-momentum tensor is given by a 
combination of a perfect-fluid and a radiated flow of particles:
\be\label{totT}
T^{\mu\nu}= T^{\mu\nu}_{\rm PF} + T^{\mu\nu}_{\rm R} 
\ee
so that $ T^{\mu\nu}_\psi= T^{\mu\nu}_{\rm PF}$ and 
$T^{\mu\nu}_{\rm ext}= T^{\mu\nu}_{\rm R}$; thus, the 
energy-momentum tensor of the perfect-fluid alone will not conserve by separate;  
 ${\cal F}_\nu$ will represent the ``forces'' of the radiated flow acting 
on the perfect fluid in form of {\it collisions}.

The energy-momentum conservation equations (\ref{divT}) can 
be written in 3+1 form as well. 
The projection of Eq. (\ref{divT}) along $n^\mu$ leads to the 
energy conservation equation,

\be\label{ECEf}
\partial_t E_\psi +  N^l\partial_l  E_\psi 
+ \frac{1}{N}\,^3\nabla_l\left(N^2 J^l_\psi\right) = 
N \left( S^{ij}_\psi K_{ij} +   E_\psi K\right)  +  N n_\nu {\cal F}^\nu 
\,\,\,\,.
\ee
Explicitly $n_\nu {\cal F}^\nu= -N  {\cal F}^t$.

On the other hand, the projection of Eq. (\ref{divT}) on $\Sigma_t$ leads 
to the momentum conservation equation,

\be\label{ECMf}
\partial_t J_i^\psi + N^l\partial_l J_i^\psi +  
J_l^\psi \partial_i N^l + N (\,^3\nabla_l \,_\psi\,\!S^l_{\,\,\,i})
=   NK J_i^\psi - \left(\,_\psi\,\!S_{\,\,\,i}^l + E_\psi \delta_{\,\,\,i}^l
\right)\,^3\nabla_l N  - \, ^3{\cal F}_i N \,\,\,\,.
\ee
where 
\be
\,^3{\cal F}_i= h_{i\mu}{\cal F}^\mu= -N_i {\cal F}^t + 
h_{ij} {\cal F}^j \,\,\,.
\ee

\subsection{Tetrads}
In many applications the use of tetrad components of tensors 
(hereafter physical components) are better adapted to a problem than 
the coordinate components. 
This will be the case when writing the equations of relativistic transport 
for the radiated flow and the equations of motion for the matter. 

In the context of the 3+1 formalism the 
tetrad we use is the local tetrad of the Eulerian observer which is
given by $\{n^\mu, e^j_{(i)}\}$, that is, by 
the time-like vector normal to $\Sigma_t$ and by a triad on $\Sigma_t$. 
In covariant notation that tetrad is given by:
\be\label{tet}
e_{(\mu)}= q^\nu_{(\mu)}\frac{\partial}{\partial x^\nu}\,\,\,,
\ee
where $\partial/\partial x^\nu$ denotes the coordinate basis of the 
spacetime, and $q^\nu_{(\mu)}$ are the {\it tetrad coefficients} that 
allow the normalization. For instance, it turns that 
$e^\mu_{(t)}\equiv n^\mu$.

The inverse relationship of Eq. (\ref{tet}) is given by
\be
\frac{\partial}{\partial x^\mu}= e^{(\nu)}_\mu e_{(\nu)}\,\,\,\,,
\ee 
where the coefficients $e^{(\nu)}_\mu$ are related to $q^\nu_{(\mu)}$ by 
the completeness relations 
$e^{(\alpha)}_\nu q^\nu_{(\beta)}= \delta^{(\alpha)}_{\,\,\,(\beta)}$, 
and  $e^{(\alpha)}_\nu q^\mu_{(\alpha)}= \delta^{\mu}_{\,\,\,\nu}$ .

A tetrad is not uniquely defined. The Lorentz invariance SO(3,1) leaves the 
freedom on the choice of the six parameters that rotate and 
boost frames. As mentioned, a convenient choice is as follows,
\bea \label{tt}
e^{(i)}_t &:=& -e^{(i)}_j N^j \,\,\,\,,\\
\label{jauge22}
e^{(t)}_i &:=& 0  \,\,\,\,,\\
 \label{ttN}
e^{(t)}_t &=& N \,\,\,\,,\\
\label{deftriad}
h_{ij} &=& e^{(l)}_i e^{(l)}_j  \,\,\,\,. 
\eea

The inverse relations are,
\bea \label{ttti}
q^{i}_{(t)} &=& \frac{N^i}{N}  \,\,\,\,,\\
 \label{ttt2}
q^t_{(i)} &=& 0\,\,\,\,,\\
 \label{ttt3}
q^{t}_{(t)} &=& N^{-1} \,\,\,\,,\\
\label{deftriaddual}
h^{ij}  &=& q^{i}_{(l)} q^{j}_{(l)} \,\,\,\,. 
\eea
here $e^{(l)}_i q^{j}_{(l)}= \delta^i_{\,\,\,j}$ 
$e^{(i)}_l q^{l}_{(j)}= \delta^{(i)}_{\,\,\,(j)}$ .

This choice of a tetrad is compatible with the 3+1 
decomposition of the four-metric (\ref{3+1metric}), so 
that
\be
ds^2= e^{(\alpha)}_\mu e^{(\beta)}_\nu \eta_{(\alpha)(\beta)} dx^\mu dx^\nu\,\,\,,
\ee
with $e^{(\alpha)}_\nu$ given by Eqs. (\ref{tt})$-$(\ref{deftriad}) and 
$\eta_{(\alpha)(\beta)}$ stands for the Minkowski metric. 

Trivial Lorentz transformations of the chosen tetrad 
relates the Eulerian observer with other 
possible frames. Moreover, 
the best choice for a triad $e^{(l)}_i$ on $\Sigma_t$ will depend on 
the particular coordinates used on the hypersurface $\Sigma_t$ 
(see Sec. VI for the case of spherical symmetry).

Finally, the transformation law for the components of tensors tangent to $\Sigma_t$ 
from the coordinate base to the triad is as follows,
\bea
N^{(i)} &=& e^{(i)}_l N^l\,\,\,,\\
J_{(i)} &=& q_{(i)}^l J_l\,\,\,,\\
K^{(i)}_{\,\,\,\,(j)} &=& e^{(i)}_l q_{(i)}^m K^{l}_{\,\,\,\,m}\,\,\,,\\
S^{(i)}_{\,\,\,\,(j)} &=& e^{(i)}_l q_{(i)}^m S^{l}_{\,\,\,\,m}\,\,\,.
\eea
The inverse relationships are obtained from above in the obvious way. 
The triad indices [i.e., spatial indices within `()'] 
are raised and lowered with 
$\delta^{(i)}_{\,\,\,(j)}$ (i.e., the triad-covariant and 
triad-contravariant components of 3-tensors are identical to each other). 
Four-tensor components transform in a similar way using the four-dimensional 
tetrad coefficients and $\eta_{(\mu)(\nu)}$ (resp. $\eta^{(\mu)(\nu)}$) 
to lower (resp. raise) indices.

The use of the tetrad formalism will be useful 
to recast the 3+1 matter equations and later on 
to write the relativistic Boltzmann equation in a useful manner. 
For instance Eq. (\ref{ECEf}) reads,
\be \label{ECEftet}
 \partial_{(t)}E_\psi + \,^3\nabla_{(i)}J^{(i)}_\psi - E_\psi K +  2J^{(i)}_\psi 
 \,^3\nabla_{(i)}\nu  -\,_\psi\,\!S^{(i)(j)} K_{(i)(j)} = -{\cal F}^{(t)} \,\,\,\,.
\ee
where $\partial_{(t)}= n^\mu\partial_\mu$.

On the other hand, the momentum conservation equation (\ref{ECMf}) can be written as
\be \label{ECMftet}
\partial_{(t)} J^{(i)}_\psi +\,^3\nabla_{(j)} S^{(i)(j)}_\psi
+ \left[ S^{(i)(j)}_\psi + E_\psi \delta^{(i)(j)}\right] \,^3\nabla_{(j)}\nu - 
J^{(i)}_\psi K 
+ J^{(l)}_\psi\, \left( {\cal O}^{(i)}_{(t)(l)} - K^{(i)}_{\,\,\,(l)}\right)
 = -\,^3{\cal F}^{(i)} \,\,\,\,.
\ee

We remind that the covariant-derivative components with respect to a tetrad uses 
the four-Ricci rotation coefficients (RRC) as a connection. 
We define them as follows,
\be \label{RRC}
{\cal O}^{(\alpha)}_{(\beta)(\gamma)} :=
e^{(\alpha)}_\mu q^\nu_{(\beta)} \nabla_\nu q^\mu _{(\gamma)} 
=  q^\sigma _{(\beta)} e^{(\alpha)}_\mu\left(\partial_\sigma q^\mu_{(\gamma)} + 
q^\lambda_{(\gamma)} \Gamma^\mu _{\lambda \sigma}\right) \,\,\,\,.
\ee
The three-RRC have an identical expression 
by restricting the above definition to pure spatial indices and using the 
three-covariant derivative. While the above definition requires the 
Christoffel symbols, these can be avoided by using the representation of the RRC 
in terms of the {\it structure constants}  
\cite{msnotes,Chandra,msrmf}.
 
Although the 3+1 Einstein equations can be written following a tetrad approach 
\cite{msnotes}, for our purposes this will not be necessary and thus we will not 
pursue the issue here.

\section{Perfect fluids with sources}
For the specific applications we have in mind, 
a combination of a perfect-fluid and a radiated flow will be 
considered. Then in Eq. (\ref{totT}) we assume 
\be
T^{\mu\nu}_{\rm PF}= (\rho +p)u^\mu u^\nu + p g^{\mu\nu}\,\,\,\,.
\ee
For the moment the form of the radiating part $T^{\mu\nu}_{\rm R}$ is not specified. 
This will be treated in detail in Sec. IV.

The corresponding 3+1 matter variables of the fluid are,
\bea \label{EPPF}
E_{\rm PF} &=& (\rho +p)\Gamma -p \,\,\,\,,\\
\label{JPPF}
J^{(i)}_{\rm PF} &=& (E_{\rm PF} + p )\,^3\,\!U^{(i)} \,\,\,\,,\\
\label{SPPF}
S^{(i)(j)}_{\rm PF} &=& (E_{\rm PF} + p ) \,\,^3\,\!U^{(i)}\,\,^3\,\!U^{(i)} + 
\delta^{(i)(j)}p \,\,\,\,,\\
\label{SSPPF}
S_{\rm PF}&=& (E_{\rm PF} + p ) \,(\,^3\,\!U^{(i)})^2 + 3p \,\,\,\,,\\
\label{lorentz}
\Gamma &:=& -n_\mu u^\mu = u^{(t)}= \left[ 1- (\,^3\,\!U^{(i)})^2\right]^{-1/2}\,\,\,\,,
\eea
where 
\bea \label{P3U}
\,^3\,\!U^{(i)} &:=& \frac{u^{(i)}}{\Gamma} =
\frac{e^{(i)}_l}{N}\left(V^l- N^l\right) 
= \frac{1}{N}\left(V^{(l)}- N^{(l)}\right) \,\,\,\,,\\
\label{Vcoor}
V^l &:=& u^l/u^t\,\,\,\,.
\eea

The equation for conservation of energy (\ref{ECEftet}) applied to a 
perfect fluid with sources then reads \cite{msnotes,msrmf,msdgfm,gour4}
\be\label{nrjEuler}
\partial_{(t)} E + \,^3\nabla_{(l)}\left[(E+p)\,^3\,\!U^{(l)}\right]
+ (E+p) \left[ 2\,\,^3\,\!U^{(j)}a_{(j)} - 
\,^3\,\!U^{(l)}\,\, ^3\,\!U^{(j)}K_{(l)(j)} -K 
\right]  = -{\cal F}^{(t)} \,\,\,\,.
\ee
 
On the other hand, the momentum conservation equation 
(\ref{ECMftet}) applied to a perfect-fluid 
with sources can be written as an Euler equation for 
the velocity field. This reads \cite{msnotes,msrmf,msdgfm,gour4},

\bea \label{EulerRG}
& & \partial_{(t)} \,^3\,\!U^{(i)} + \,^3\,\!U^{(j)}\,\,
^3 \nabla_{(j)}\,^3\,\!U^{(i)}  = 
- \frac{1}{E_{\rm PF}+p} \left[ \,^3\partial^{(i)}p + \,^3\,\!U^{(i)} \partial_{(t)}p 
\right] - a_{(j)} \nonumber \\
\!\!\!\!\!\!\!\!\!\!\!\!
&+& \,^3\,\!U^{(i)} \,\,^3\,\!U^{(l)}\left(a_{(l)}  
- \,^3\,\!U^{(j)}K_{(l)(j)} \right) 
- \,^3\,\!U^{(l)} \left( {\cal O}^{(i)}_{(t)(l)} - K^{(i)}_{\,\,\,(l)}\right)
  + \frac{1}{E_{\rm PF}+p}\left(\,^3\,\!U^{(i)}{\cal F}^{(t)} - {\cal F}^{(i)}\right)
\,\,\,\,.
\eea
where 
\be\label{euleractet}
a_{(i)}=\,^3\nabla_{(i)}[{\rm ln}N].
\ee
are the physical components of Eq.(\ref{eulerac}). This Euler equation 
is the version in physical components 
(with the extra dissipative terms ${\cal F}^{(\mu)}$) of Eq. (2.29) of Ref.\cite{gour4} 
(see also Ref. \cite{warn}).

\section{Relativistic transport theory}
\bigskip
We shall consider the phenomenon of relativistic transport of 
massive and massless particles (hereafter radiation) within a dense medium. 
We shall follow closely the formalism developed by Lindquist 
\cite{Lind}, but we will not repeat the details here. 
The radiated particles of the specie `${\rm R}$' will be treated classically as 
point particles except when interacting with the dense medium. The interactions 
and its quantum mechanical effects will be 
ultimately translated as emission rates and opacity functions. The particles
will be characterized thus by a four-momentum 
\be
p^\mu= \frac{dx^\mu}{d\lambda}\,\,\,,
\ee
where $d\lambda$ corresponds to an affine parameter (for massless particles) 
or to the proper-time per mass unit (for massive particles).

According to this formalism, one postulates the existence of 
a scalar function $F_{\rm R}(x^\mu,p^\mu)$ (the {\it invariant} distribution function 
for the specie ${\rm R}$) which is a function from the phase space 
coordinates $(x^\mu, p^\mu)$ to the reals. 
Actually, since we will be interested in particles 
satisfying the mass shell condition
\be\label{curvms}
g_{\mu\nu}p^\mu p^\nu = -\tilde m\,\,\,,
\ee
where $\tilde m= 0,1$ for massless or massive particles respectively, 
$F_{\rm R}$ will be a function defined on the {\it reduced} phase-space.

The invariant distribution function so introduced will be such that the 
number density four-vector and the 
energy-momentum tensor of the particles are respectively \cite{Lind},

\be \label{partflux}
j^{\mu}_{\rm R} = \int p^\mu F_{\rm R}(x^\mu,p^\mu)dP \,\,\,\,.
\ee

\be \label{Tpart}
T^{\mu \nu}_{\rm R} = \int p^\mu p^\nu F_{\rm R}(x^\mu,p^\mu)dP \,\,\,\,.
\ee
where $dP$ is the invariant volume of the momenta space on shell 
\cite{Lind}:

\be\label{dPfc}
dP =  - N \sqrt{h}\frac{d^3p}{p_t}\,\,\,\,. 
\ee 
where $d^3p:= dp^1 dp^2 dp^3$ represents a coordinate 3-volume element 
and $h$ is the determinant of the 3-metric.

The distribution function $F_{\rm R}$ is related to the dimensionless distribution 
function $\bar F_{\rm R}$ by
\be
F_{\rm R}=  \frac{ g_{\rm R}}{ 8\pi^3 \hbar^3_{\rm Pl}} \bar F_{\rm R} \,\,\,\,.
\ee
where $g_{\rm R}$ is the statistical weight of the particles of the 
specie ${\rm R}$ (e.g., $g=1,2$ for neutrinos and photons respectively).

As in the case of the perfect fluid, it is useful to refer the 
components of four-momenta to a tetrad. In the context of the 3+1 formalism we 
have,
\bea
e &:=& p^{(t)}= N p^t \,\,\,\,,\\
p^{(i)} &=& e^{(i)}_l (p^l - p^t N^l) \,\,\,\,,
\eea
with the inverse relationships given by
\bea 
p^t &=& e/N \,\,\,\,,\\
\label{p^i}
p^i &=&  e\frac{N^i}{N} +  q_{(l)}^i p^{(l)} \,\,\,\,,
\eea
where $p^{(i)}$ are the physical spatial  
components of the 4-momentum (i.e. the spatial physical components of the
 projection of $p^\mu$ onto 
$\Sigma_t$: $^3\,p^\mu:= h^\mu_{\,\,\,\nu}p^\nu$). 
The ratio $p^{(i)}/e$ corresponds to the local velocity of particles 
measured by ${\cal O}_{\rm E}$.

Introducing the magnitude of the three-momentum as,
\be
p^2 := p_{(i)}p^{(i)}\,\,\,,
\ee
it is easy to see that (\ref{curvms}) simply becomes 
\be \label{curvms2}
e^2 = p^2 + \tilde m^2 \,\,\,\,.
\ee
Here $e$ is the energy (per mass-unit in the case of massive particles)
as measured by the Eulerian observer. Therefore, from Eq. (\ref{dPfc}), 
one obtains,
\be\label{dPfc2}
dP =   \sqrt{h}\frac{d^3p}{e\left(1 + 
\frac{p_{(l)} e^{(l)}_i N^i}{EN}\right)}\,\,\,\,. 
\ee 
where we used that $p_t=  e^{(\mu)}_t p_{(\mu)}= Np_{(t)}- p_{(l)} e^{(l)}_i N^i= 
-Ne - p_{(l)} e^{(l)}_i N^i$.

When changing variables in the momentum space using (\ref{p^i}) 
and (\ref{curvms2}) a straightforward manipulations show that
\be
d^3p = \frac{dp^{(1)}dp^{(2)}dp^{(3)}}{\sqrt{h}}
\left(1 + \frac{p_{(l)} e^{(l)}_i N^i}{eN}\right)\,\,\,\,.
\ee
where it was used the fact that ${\rm det}[q_{(l)}^i]= 1/\sqrt{h}$. 
Then finally
\be\label{dPfc3}
dP =   \frac{dp^{(1)}dp^{(2)}dp^{(3)}}{e}\,\,\,\,. 
\ee 
which has exactly the same form of flat spacetimes. 

The use of physical spherical variables in momentum space leads to 
\be \label{dPsph}
dP = p^2dp d\Omega_p/e = p de  d\Omega_p  \,\,\,\,.
\ee
Indeed, this is the useful expression when dealing with spherical symmetry.

The use of tetrad components allows us to write Eqs. (\ref{partflux}) and 
(\ref{Tpart}), as follows,

\be \label{partfluxtet}
j^{(\mu)}_{\rm R} = \int p^{(\mu)} F_{\rm R}(x^\lambda, p^{(\lambda)})dP \,\,\,\,,
\ee
\be \label{Tparttet}
T^{(\mu) (\nu)}_{\rm R} = \int p^{(\mu)} p^{(\nu)} F_{\rm R}(x^\lambda, p^{(\lambda)}) dP \,\,\,\,.
\ee
Therefore, the corresponding 3+1 matter variables are
\bea\label{EPk}
E_{\rm R} &=& \int e^2  F_{\rm R}(x^\lambda, p^{(\lambda)}) dP \,\,\,\,,\\
\label{JPk}
J^{(i)}_{\rm R} &=& \int  e p^{(i)}  F_{\rm R}(x^\lambda, p^{(\lambda)}) dP \,\,\,\,,\\
\label{SPk}
S^{(i)(j)}_{\rm R} &=& \int  p^{(i)} p^{(j)}  F_{\rm R}(x^\lambda, p^{(\lambda)}) dP \,\,\,\,,\\
\label{SSPk}
S_{\rm R} &=& \int p^2 F_{\rm R}(x^\lambda, p^{(\lambda)}) dP \,\,\,\,.
\eea

According to (\ref{3+1mattvartot}) the total 3+1 matter variables are,
\bea \label{EPPFT}
E &=& E_{\rm PF} + E_{\rm R}\,\,\,,\\
\label{JPPFtot}
J^{(i)} &=& J^{(i)}_{\rm PF} + J^{(i)}_{\rm R}  \,\,\,\,,\\
\label{SPPFT}
S^{(i)(j)} &=& S^{(i)(j)}_{\rm PF} + S^{(i)(j)}_{\rm R}  \,\,\,\,,\\
\label{trSPPFT}
S &=& S_{\rm PF} + S_{\rm R} \,\,\,\,.
\eea
where we remind that the quantities labeled with `PF' are given by Eqs. 
(\ref{EPPF})$-$(\ref{SSPPF}). It is understood in these expressions that 
the sum of the different quantities extend to all the species ${\rm R}$ 
considered.

\bigskip
\subsection{The Boltzmann equation in curved space-times}
\bigskip

We defined a ``macroscopic'' 4-current density number of particles of the 
specie ${\rm R}$ by 
Eq. (\ref{partflux}). 
The number density of particles as measured in the local 
frame of an observer ${\cal O}_p$ with four-velocity $v^\mu$) is
\be
 n_p^{\rm R} := - v_\mu j^\mu_{\rm R} 
= - \int v_\mu p^\mu  F_{\rm R} (x^\lambda, p^{(\lambda)}) dP \,\,\,\,.
\ee
Therefore, the number of particles $dN_{\rm R}$ with momenta between 
$p^\mu$ and $p^\mu + dp^\mu$ 
crossing the volume element $dV$ of the space-like hypersurfaces 
orthogonal to $v^\mu$ and which is centered at some point $x^\mu$ of spacetime is
\be
dN_{\rm R}= F_{\rm R} (x^\lambda, p^{(\lambda)}) \left(- v_\mu p^\mu\right)
dVdP\,\,\,\,\,.
\ee
The quantity $dW= \left(- v_\mu p^\mu\right) dVd\lambda$ 
represents the four-volume spanned by the flow of particles (world lines) crossing 
$dV$, which is given by the element of hypersurface $dV$ with 
normal $v^\mu$ and the particle's 
infinitesimal displacement orthogonal to $dV$ given by  
$dl= v_\mu p^\mu d\lambda$ \cite{Lind}. The quantity 
$\left(- v_\mu p^\mu\right) dV$ is in fact 
the correct Lorentz invariant four-volume element. From the relativistic form 
of Liouville's Theorem (see Ref. \cite{Lind} for the details) 
$dW dP$ remains invariant along a given set of trajectories. 
Therefore, the change in the number of world lines within 
$dW dP$ is proportional to the change in $F_{\rm R}$
\bea
\delta \left(dN_{\rm R}\right) &=& \left[ \frac{\partial F_{\rm R}}{\partial x^\alpha} 
dx^\alpha +  \frac{\partial F_{\rm R}}{\partial p^\alpha} 
dp^\alpha\right]\left(- v_\mu p^\mu \right) dV dP 
\nonumber \\
&=& \left[ \frac{\partial F_{\rm R}}{\partial x^\alpha} 
p^\alpha +  \frac{dp^\alpha}{d\lambda} \frac{\partial F_{\rm R}}{\partial p^\alpha} 
 \right] dW dP \,\,\,\,\,.
\eea
 The evolution for 
$p^\mu$ will be thus governed by the equations of motion of individual 
particles:
\be \label{motioneq}
\frac{dp^\mu}{d\lambda} = - p^\sigma p^\nu \Gamma^{\mu}_{\sigma \nu} 
+ {\cal F}_{\rm fields}^\mu
+ {\cal F}_{\rm coll}^\mu
\,\,\,\,.
\ee

This equations shows that the acceleration of the particles is due to 
1) the space-time curvature 2) the forces arising from the 
interaction of the particles with fundamental fields 
other than the gravitational one, 3) the interaction 
with other particles that can be represented by ``collisions''. 
For our purposes, we will consider that ${\cal F}_{\rm fields}^\mu=0$. 
That is to say, the only fundamental field we consider is the gravitational 
one. All other interactions like the weak ones (in the case of neutrinos) 
will be treated phenomenologically as collision terms, and therefore, 
the set of equations will not include gauge-fields, but rather 
involve macroscopic quantities that characterize the medium and 
which are obtained from field theory in a similar fashion as one 
obtains the equation of state of the matter.
 
The relativistic Boltzmann equation (RBE) then reads,
\be \label{Boltzeq}
 \hat{\mbox{\bm{$L$}}} F_{\rm R} = \left(\frac{dF_{\rm R}}{d\lambda}\right)_{\rm coll} \,\,\,\,,
\ee
where 
\be \label{Liouvop}
\hat{\mbox{\bm{$L$}}} := p^\mu\frac{\partial }{\partial x^\mu} - 
p^\nu p^\sigma \Gamma^\mu _{\nu \sigma}
 \frac{\partial }{\partial p^\mu} \,\,\,\,,
\ee
is the relativistic Liouville operator often written by the fuzzy 
notation $p^\alpha D/dx^\alpha$ (the directional derivative of $F_{\rm R}$ along 
the phase flow), and
$\left(\frac{dF_{\rm R}}{d\lambda}\right)_{\rm coll}= -{\cal F}_{\rm R\,\, coll}^\alpha
\frac{\partial F_{\rm R}}{\partial p^\alpha}$ 
represents collectively the scattering, absorption and emission 
processes between the particles of the specie ${\rm R}$ and the medium. 
In the absence of collisions, the distribution function remain 
constant along the particle's path (i.e., along particle's geodesics). 
In the language of differential geometry, the operator 
$D/dx^\alpha= 
\frac{\partial }{\partial x^\alpha} - 
 p^\lambda \Gamma^\mu _{\alpha \lambda}
 \frac{\partial }{\partial p^\mu}$ corresponds to a coordinate 
basis vector of the horizontal part of the  
tangent space of the bundle ${\cal B}$ over the spacetime $M^4$ \cite{Lind}. 
It is important to emphasize that the mass shell condition 
implies that the distribution function will not be defined 
over the entire tangent space but only on that part where  
$p_\mu p^\mu = -\tilde m^2 $, that is the distribution function 
will be defined only on the {\it sphere bundle} (the 
subbundle of tangent vectors of fixed length). One can incorporate 
this restriction in the Liouville operator by treating the spatial 
part $p^i$ of the four-momenta as independent components. Then 
\be
\left(D/dx^\alpha\right)_{\rm ms} = \frac{\partial F_{\rm R}}{\partial x^\alpha} - 
p^\lambda  \Gamma^i _{\alpha \lambda}
 \frac{\partial F_{\rm R}}{\partial p^i}\,\,\,.
\ee 

\bigskip
\subsection{Tetrad representation and the 3+1 RBE}
\bigskip 
It will be convenient to use a tetrad components to re-write the 
RBE. By employing the tetrad formalism it's easy to show that the RBE reads
\be \label{Boltzeq4}
\left(\,p^{(\alpha)}q^{\mu}_{(\alpha)}\frac{\partial}{\partial x^{\mu}} - 
p^{(\beta)}p^{(\alpha)}{\cal O}^{(\delta)}_{(\beta)(\alpha)}
\frac{\partial}{\partial p^{(\delta)}}\,\right)
F_{\rm R}(x^{\alpha},p^{(\sigma)}) = \left(\frac{d F}{d\lambda}\right)_{\rm coll}
\,\,\,\,,
\ee

Let us consider Eq.(\ref{Boltzeq4}), and split it in terms of 
temporal and spatial contributions. First we define,
\be
L_p := p^{(b)}p^{(a)}{\cal O}^{(d)}_{(b)(a)}\frac{\partial}{\partial p^{(d)}}
=p^{(b)}p^{(a)}{\cal O}^{(t)}_{(b)(a)}\frac{\partial}{\partial e}
+ p^{(b)}p^{(a)}{\cal O}^{(i)}_{(b)(a)}\frac{\partial}{\partial p^{(i)}}
\,\,\,\,,
\ee
where the notation $e= p^{(t)}$ was used. 
The properties of the RRC and some 
straightforward calculations show a useful relationship between 
the four-RRC and the physical components of 3+1 variables. For instance:
\bea
{\cal O}^{(t)}_{(t)(i)} &=& {\cal O}^{(i)}_{(t)(t)}= a_{(i)}\,\,\,\,,
\nonumber \\
{\cal O}^{(t)}_{(i)(j)} &=& {\cal O}^{(i)}_{(j)(t)}= -K_{(i)(j)}\,\,\,\,,
\nonumber \\
{\cal O}^{(\mu)}_{(t)(\mu)} &=& {\cal O}^{(\mu)}_{(i)(\mu)} = 0 
\,\,\,({\rm no\,\,sumation\,\,on}\,\,\mu) \nonumber \\
{\cal O}^{(i)}_{(t)(j)} &=& -\frac{1}{2}\left( 
-\frac{\partial_{(i)} N^{(j)}}{N} + q^l_{(m)}
\frac{N^{(m)}}{N} \partial_{(i)}e^{(j)}_{l} +  q^l_{(i)} 
\partial_{(t)}e^{(j)}_{l} \,\,-\,\,(i)\longleftrightarrow \,\,(j)\right)  \nonumber \\
{\cal O}^{(l)}_{(i)(j)} &=& ^3\,\!{\cal O}^{(l)}_{(i)(j)}
\eea
 Here
 $^3\,\!{\cal O}^{(l)}_{(i)(j)}$ are the 3-Ricci rotation coefficients, 
i.e., the RRC associated to the local basis frame on $\Sigma_t$, and 
$\partial_{(t)}= \frac{1}{N}\frac{\partial}{\partial t} + 
\frac{N^{(i)}} {N} \partial_{(i)}$.

In this way we obtain
\bea
L_p= \left(  e p^{(i)} a_{(i)} -  p^{(i)}p^{(j)}K_{(i)(j)}\right)
\frac{\partial}{\partial e}
+ \left( e^2 a^{(i)} + e p^{(i)} {\cal O}^{(i)}_{(t)(j)} - 
e p^{(j)} K^{(i)}_{\,\,\,\,(j)} + p^{(l)} p^{(j)}\,^3\,\!{\cal O}^{(l)}_{(i)(j)}\right)\frac{\partial}{\partial p^{(i)}}
\eea

Finally, the 3+1 decomposition of Eq.(\ref{Boltzeq4}) is, 
\bea \label{3+1RBE1}
& &\left[\, e \partial_{(t)}
 + p^{(i)} \partial_{(i)} 
- \left( e p^{(i)} a_{(i)} -p^{(i)}p^{(j)} K_{(i)(j)}\right)
\frac{\partial}{\partial e}\right. \nonumber \\
&&
\left. 
-\left(e^2a^{(l)} + e p^{(i)} {\cal O}^{(i)}_{(t)(j)} - 
e p^{(j)} K^{(i)}_{\,\,\,\,(j)}  + p^{(i)}p^{(j)}\, ^3\,\!
{\cal O}^{(l)}_{(i)(j)}
\right) \frac{\partial}{\partial p^{(l)}} \,\right]
F_{\rm R}(x^{\alpha},p^{(c)}) = \left(\frac{d F}{d\lambda}\right)_{\rm coll}
\,\,\,\,.
\eea
It is to be noted that the properties of RRC imply that 
$ ^3\,\!{\cal O}^{(l)}_{(i)(l)}$ (no sum convention) are null. 
Therefore, the above equation can be further simplified.

Equation (\ref{3+1RBE1}) is the 3+1 version of the 
of RBE, here written in terms of physical components. 
The mass shell condition $e^2= p^{(i)}p_{(i)} + \tilde m^2$ can be 
imposed on the RBE by considering, for instance, $p^{(i)}$ 
as independent variables. In that case 
$F_{\rm R}$ is to be considered as though it does not depend explicitly 
on $e$, i.e., $\partial F_{\rm R}/\partial e =0$. Alternatively, the use 
of spherical-like variables in momentum space (see next section) 
will allow us to consider $e$ as independent variable and 
$p^2= p^{(i)}p_{(i)}$ as the dependent one.

\bigskip
\subsection{Collisions}
\bigskip

As we stressed before, particles may be submitted to 
collision forces arising from the interacting medium 
(in the case of neutrinos these forces 
come from weak interactions with baryons). 
Let us remind that the collision integral as conceived 
originally by Boltzmann assumes that 
 the interacting medium has a known distribution function. 
That's it, the distribution function of the medium is a data 
of the problem.

In the present case, the interacting medium is to be considered not 
as particles but rather as a 
fluid field, namely a perfect fluid. 
Thus, for our purpose, it will be more convenient 
to characterize the collision integral in terms of scalar 
functions as it is usual in transport theory. In this way collisions 
 will be represented macroscopically by the so-called 
{\it invariant opacity} $o(x^\mu,p^\mu)$ and the invariant emissivity 
$\Upsilon(x^\mu,p^\mu)$.

We then assume that the collision integral takes the following form
\be
\left(\frac{d F_{\rm R}}{d\lambda}\right)_{\rm coll}= \Upsilon - oF_{\rm R}\,\,\,.
\ee

In terms of quantities measured in the same frame one can write
\be
\Upsilon= \frac{\eta(x^\mu,p^\mu)}{e^2}= \frac{\eta'(x'^\mu,p'^\mu)}{e'^2} \,\,\,,
\ee
here $\eta$ and $e$ being the matter emissivity and the particle energy 
respectively, both measured in the same frame. In the same way, one can 
introduce the matter opacity as
\be\label{invopac}
o= e\chi(x^\mu,p^\mu)= e'\chi'(x'^\mu,p'^\mu) \,\,\,\,,
\ee
where $\chi \sim 1/\ell$, $\ell$, being the mean free path of the 
particle in the corresponding frame. 

The collision term takes then the useful form
\be
\left(\frac{d F_{\rm R}}{d\lambda}\right)_{\rm coll}= e\left(\frac{\eta}{e^3} 
- \chi F\right)\,\,\,.
\ee

The opacity $\chi$ and the absorption coefficient $\kappa$ are related 
by
\be
\chi= \kappa n\,\,\,,
\ee
where $n$ is the proper number density of particles that composes the medium 
(e.g., the baryon density), such that $n:= -j^\mu u_\mu$ where $j^\mu$ is the 
four-current of baryons.

In this way, an alternative form of the collision integral is
\be
\left(\frac{d F_{\rm R}}{d\lambda}\right)_{\rm coll}= \kappa_e n\left({\cal S}- 
F_{\rm R}\right)  \,\,\,\,,
\ee
where $k_e= e\kappa$, and ${\cal S}= \Upsilon/o$ is usually referred to as 
the {\it effective} source function.

It is to be emphasized that quantities measured in different frames 
are related to each other via the invariant quantities and Lorentz transformations, 
for instance, 
the relationship between the opacities meaured in the Eulerian frame 
and those of the proper frame of the fluid are given, according to Eq. (\ref{invopac}), 
by
\be
e\chi = e_p \chi_p  \,\,\,\,,
\ee 
where quantities in the left-hand-side (l.h.s) refer to the Eulerian 
frame, while the quantities in the right-hand-side (r.h.s) refer to the proper 
frame of the fluid.
Since physical components of four vectors in both frames are related 
by a Lorentz transformation, for instance
\be
p^{(\mu)}_p= \Lambda^{(\mu)^p}_{\,\,\,(\nu)} p^{(\nu)}  \,\,\,\,,
\ee
where 
\be\label{TLH}
\left( \Lambda^{(\mu)^p}_{\,\,\,(\nu)} \right)\,\,\,= \,\,
\left( \begin{array}{cc} 
\Gamma &  -\,^3\,\!U^{(i)}\Gamma \\
  -\,^3\,\!U^{(i)}\Gamma  & \delta^{(i)}_{(j)} + \,^3\,\!U^{(i)} 
\,^3\,\!U^{(j)}\frac{\Gamma^2}{\Gamma +1} \\
\end {array}
\right)\,\,\,\,\,\,,
\ee
with $ \,^3\,\!U^{(i)}$ given by Eq.(\ref{P3U}) and $\Gamma$ by Eq.(\ref{lorentz}), 
then for the time components,
\be
e_p= \Gamma \left(e - \,^3\,\!U^{(i)}  p^{(i)} \right)= 
e \Gamma \left(1 - \,^3\,\!U^{(i)} v^{(i)} \right) =
e \Gamma \left(1 - ||\,^3\,\!U^{(i)}|| \,||v^{(i)}|| \cos(\theta_{\rm R})\right)  \,\,\,\,,
\ee
where $v^{(i)}:= p^{(i)}/e$ represents the velocity of the 
particles with respect to the Eulerian frame. In the last formula 
one recognizes the well known formula for the energy shift due to the relative 
motion of observers. The type of shift (red or blue) will depend on the 
angle $\theta_{\rm R}$ between the propagation vector of the particles 
$v^{(i)}$ and the velocity 
of the fluid $\,^3\,\!U^{(i)}$ (i.e., blue or red shift if 
the fluid is approaching or receeding respectively from the Eulerian observer). 
Therefore the transformation formula between opacities yields
\be
\chi= \chi_p \Gamma \left(1 -  ||\,^3\,\!U^{(i)}|| \,||v^{(i)}|| \cos(\theta_{\rm R}) \right)
\,\,\,\,.
\ee
In the case of massless particles $||v^{(i)}||=1$. In a similar way one can obtain 
the transformation formulae for the absortion coefficients and the emissivities.

\bigskip
\subsection{Conservation equations for the radiated flow}
\bigskip
The particle number current and 
the energy-momentum tensor of the radiated particles was 
introduced by Eqs. (\ref{partflux}) and (\ref{Tpart}) respectively. 
For instance, in the case of perfect quantum gases in thermal equilibrium
(i.e., Fermi and Bose gases) the above definitions 
allows one to recover the usual macroscopic 
expressions for the energy-density, density-number and pressure parameterized 
by the temperature, particle-mass and chemical potential of the species 
as measured in the local frame.

When collisions are present, both 
the particle number and the energy-momentum tensor 
of the particles will not conserve alone since there will be exchange of 
energy and momentum with the interacting dense fluid. Thus we can expect 
that the conservation equations derived from Eqs. (\ref{partflux}) and 
(\ref{Tpart}) will have sources arising from the collision integral: 

\be
\label{consN}
\nabla_\nu j^{\nu}_{\rm R}= 
\int \
 \left(\frac{d F}{d\lambda}\right)_{\rm coll} dP   \,\,\,\,,
\ee 
and
 \be \label{conseqs3}
\nabla_\nu T^{\mu \nu}_{\rm R}
= \int \
 p^\mu \left(\frac{d F}{d\lambda}\right)_{\rm coll} dP  \,\,\,\,.
\ee 
 Then we write
\bea
\label{consN2}
\nabla_\nu j^{\nu}_{\rm R} &=& {\cal R}_{\rm R} \,\,\,,\\
 \label{conseqs4}
\nabla_\nu T^{\mu \nu}_{\rm R} &=& -{\cal F}^\mu_{\rm R}= nw^\mu_{\rm R} \,\,\,\,,
\eea
where
\bea\label{reacrate}
{\cal R}_{\rm R} &:=& 
\int \kappa (x^\mu,p^\mu)\left[{\cal S}(x^\mu,p^\mu) 
- F_{\rm R}(x^\mu,p^\mu)\right]dP \,\,\,\,, \\
 \label{forcesour}
w^\mu_{\rm R} &:=& \int p^\mu \kappa (x^\mu,p^\mu)\left[{\cal S}(x^\mu,p^\mu) 
- F_{\rm R}(x^\mu,p^\mu)\right]dP \,\,\,\,.
\eea
One can define the {\it mean} emissivity (energy/(volume$\times$time) in the fluid frame as 
\be
{\cal D}:= - u_\mu \nabla_\nu T^{\mu \nu}_{\rm R}= - nu_\mu w^\mu \,\,\,.
\ee
Since we have been using quantities measured in the Eulerian frame, 
in the above expressions we are to use the corresponding quantities with respect 
to the same observer. Namely, the particle number density 
 $n_{\rm E}$ as measured by the Eulerian observer is related to 
the proper number density of the perfect fluid $n$ by 
$n_{\rm E}= n\Gamma$. Same considerations apply for the remaining collision variables. 
For instance, the emissivity  measured 
in the Eulerian frame ${\cal D}_E= -n_\mu n w^\mu= n w^{(t)}$ is related to the 
proper emissivity ${\cal D}= -u_\mu n w^\mu$ by $ {\cal D}= \Gamma  {\cal D}_E  
\left( 1 - \,\,^3\,\!U^{(i)} \,\,^3\,\!W^{(i)}\right)$ where 
$\,^3\,\!W^{(i)}:= w^{(i)}/w^{(t)}$.

\section{Thermodynamics}
In this section the thermodynamical description of the dense matter with 
which the radiative particles interact will be presented taking into account 
the mean quantities introduced in the previous section. Such a description 
is performed in the proper frame of the fluid. We then 
assume that the equation of state of the dense matter 
(i.e., the perfect fluid) is given in parametrized 
form as follows
\bea\label{eos1}
\rho &=& \rho (s,n_1,\dots,n_m) \,\,\,\,,\\
\label{eos2}
p &=& p (s,n_1,\dots,n_m) \,\,\,\,,
\eea
where $s$ is the entropy density and $n_M$ ($1\leq M \leq m$) 
is the number density of particles 
of the specie $M$ (e.g. baryons and the different lepton flavors), 
all of them measured in the fluid frame. For instance, in the 
case of a dense matter in hydrostatic 
equilibrium composed by a mixture of neutrons, protons and electrons, the 
electron density is obtained directly from the proper baryon density $n$ 
by demanding charge neutrality and chemical equilibrium $\mu_{\rm n}= 
\mu_{\rm p} + \mu_{\rm e}$. This last condition arising from the equilibirum 
of the nuclear reactions: $n\rightleftharpoons  p + e^-$ \cite{gour2}. In that case the equation 
of state depends on $n$ solely.

Now, the equations (\ref{eos1}) and (\ref{eos2}) 
are not independent from each other but 
linked through the first principle of thermodynamics
\be\label{1stPT}
d{\rm U}= \theta d{\rm S} -p dV + \mu^{\rm R} d{\rm N}^{\rm R}
\ee 
where $\Theta$ and $\mu^{\rm R}$ are the temperature and the chemical potential 
of the specie ${\rm R}$ of the particles composing the fluid 
respectively, defined by
\bea
\Theta &=& \left(\frac{\partial\rho}{\partial s}\right)_{n_{\rm R}}\,\,\,\,,\\
\mu^{\rm R} &=& \left(\frac{\partial\rho}{\partial n_{\rm R}}\right)_{s,n_B\neq n_{\rm R}}\,\,\,\,.
\eea
Using these definitions, Eq. (\ref{1stPT}) takes the following form in terms 
of densitized quantities \cite{gour2},
\be
p= \Theta s + \mu^{\rm R} n_{\rm R} - \rho\,\,\,.
\ee
This equation is often referred to as the compatibility themodynamic condition 
between Eqs. (\ref{eos1}) and (\ref{eos2}) \cite{gour2}.

The conservation equation for the baryon number reads,
\be\label{barcons}
\nabla_\mu j^\mu =0\,\,\,\,.
\ee
where we remind that $j^\mu= nu_\mu$ is the density current of baryons and 
$n$ the proper baryon number density. 

This equation can be written explicitly as an evolution 
equation for the number density 
$n_{\rm E}:= -n_\mu j^\mu= n\Gamma$ measured by the Eulerian observer as follows

\be\label{barcons1}
\partial_t \left( \sqrt{h} n_{\rm E}\right) + \partial_i\left[\sqrt{h} 
n_{\rm E} V^i\right] =0\,\,\,\,,
\ee
where $\Gamma$ and $V^i$ are given by 
Eqs. (\ref{lorentz}) and (\ref{Vcoor}) respectively. Introducing the 
physical components of the fluid velocity field given by 
Eq. (\ref{P3U}), we have the alternative expression,

\be\label{barcons2}
\partial_t \left( \sqrt{h} n_{\rm E}\right) + \partial_i\left[\sqrt{h} 
n_{\rm E} \left( N^i + Nq^i_{(j)}\,^3\,\!U^{(j)}\right)\right] =0\,\,\,\,.
\ee

The equation of conservation of baryon number leads 
to the conserved total baryon number given by
\bea
{\cal N} &=& \int_{\Sigma_t} - j^\mu n_\mu \sqrt{h} dx^1 dx^2 dx^3\\
  &=&  \int_{\Sigma_t} n \Gamma \sqrt{h} dx^1 dx^2 dx^3\,\,\,.
\eea
The integral has compact support corresponding to the volume enveloped 
by the star surface.

In a similar way, the equation for entropy conservation reads
\be\label{entcons}
\nabla_\mu \left( su^\mu\right)= -\frac{1}{\Theta}\left( \mu^{\rm R} {\cal R}_{\rm R} + 
{\cal D}\right) \,\,\,\,,
\ee
where
\be\label{partcons}
{\cal R}_{\rm R} := \nabla_\mu \left(n_{\rm R} u^\mu\right)\,\,\,,
\ee
is the rate of particle production and we remind that 
\be
{\cal D}:= - u_\nu \nabla_\mu T^{\mu\nu}_{\rm R}=  
u_\nu \nabla_\mu T^{\mu\nu}_{\rm PF}\,\,\,\,,
\ee
is the particle's mean emissivity in the fluid frame. Therefore the sources for 
the entropy generation in a perfect fluid are from the particle 
production (e.g. neutrinos). The Eq. (\ref{partcons}) is completely 
equivalent to Eq. (\ref{consN2}) which is given in terms of the 
distribution function.

One can define the entropy per baryon $\sigma=s/n$ and use Eqs. 
(\ref{barcons}) and (\ref{entcons}) to obtain
\be\label{entcons2}
u^\mu \nabla_\mu \sigma= -\frac{1}{n\Theta}\left( \mu^{\rm R} {\cal R}_{\rm R} 
+{\cal D} \right)  \,\,\,.
\ee
Explicitly this provides an evolution equation for $\sigma$:
\be\label{entconscoor}
\partial_t \sigma + V^i \partial_i \sigma= 
 -\frac{N}{n\Theta\Gamma}\left( \mu^{\rm R} {\cal R}_{\rm R} + {\cal D}\right) \,\,\,.
\ee

Moreover, using Eqs. (\ref{lorentz}), (\ref{P3U}) and the tetrad approach  
of Sec. IIA, the Eq. (\ref{entcons2}) takes the alternative form:
\be\label{entcons3}
n^\mu \partial_\mu \sigma + \,^3\,\!U^{(i)}\partial_{(i)}\sigma
= -\frac{1}{n\Gamma \Theta}\left( \mu^{\rm R} {\cal R}_{\rm R} + {\cal D}\right)\,\,\,.
\ee
In the same way, we can introduce the particle number per baryon 
$x_{\rm R}= n_{\rm R}/n$ and write (\ref{partcons}) as
\be\label{partcons2}
u^\mu \nabla_\mu x_{\rm R} = \frac{1}{n}{\cal R}_{\rm R}\,\,\,,
\ee
which provides an evolution equation for $x_{\rm R}$:
\be\label{partconscoor}
\partial_t  x_{\rm R} + V^i \partial_i  x_{\rm R} = \frac{N}{n\Gamma}{\cal R}_{\rm R}\,\,\,,
\ee
or alternatively,
\be\label{partcons3}
n^\mu \partial_\mu x_{\rm R} + \,^3\,\!U^{(i)}\partial_{(i)}x_{\rm R} 
= \frac{1}{n\Gamma}{\cal R}_{\rm R}\,\,\,.
\ee

In order to close the whole system of equations presented so far one needs the input of 
particle physics. That is, the equation of state for nuclear matter, 
the rate of particle production and the opacities (see Ref. \cite{ST83} 
for a review). In the case of 
neutrinos emitted by nuclear matter out of beta equilibrium 
via direct and inverse $\beta$ processes during neutron 
star collapse, the rate of particle 
production, the emissivities and the opacities can be given 
in terms of rather simple formulae \cite{gour3,Haensel92} 
(see also Ref. \cite{DSI} for neutrino emissivities from quark matter in 
$\beta-$equilibrium within neutron stars and Ref. \cite{SSS} for neutrino 
emission from hot and dense atmospheres). For the 
case of reaction rates and opacities in 
Type II supernovae see for instance Ref. \cite{MWS}.

\section{Spherical symmetry}
In this and the following sections we will focus on spherically symmetric spacetimes. 
The most general line-element for such spacetimes 
according with the 3+1 decomposition of the metric Eq. (\ref{3+1metric}), writes
\be\label{ds}
ds^2= -(N^2 - N^rN_r)dt^2 - 2N_r dt dr + A^2 dr^2 + B^2(r^2 d\theta^2 
+ r^2 \sin^2 \theta d\phi^2)\,\,\,\,,
\ee
where all the metric potentials are functions of the coordinates 
$r$ and $t$ solely. The three-metric $h_{ij}$ is easily read-off from (\ref{ds}).

On the other hand, the triad coefficients are
\be \label{e}
e^{(i)}_j = \,\,{\rm diag}\left( A(r,t), 
r\,B(r,t),r\,\sin\theta\,B(r,t)\right) \,\,\,.
 \ee 
The inverse coefficients $q^i_{(j)}$ can be obtained trivially form (\ref{e}).

The extrinsic curvature can be computed from (\ref{K_ij}). We find \cite{3+1math},

\be \label{K_ijsph}
\left(K_{ij}\right)\,\,=\,\,  \left(
\matrix{ -{\frac{A}{N}}\left(\partial_t A 
       + \partial_r N^{(r)}\right)
   & 0 & 0 \cr 0 & -\frac{rB^2}{N}\left(
\frac{N^{(r)}}{A} + \frac{r}{B}\,\partial_t B + 
   \frac{r\,N^{(r)}\,\partial_r B}{AB} 
\right) & 0 \cr 0 & 0 & 
 -\frac{rB^2\sin^2(\theta)}{N}\left(
  \frac{N^{(r)}}{A} +
  \frac{r}{B}\,\partial_t B + 
  \frac{r\,N^{(r)}\,\partial_r B}{AB}\right)
      \cr  }   \right)
\,\,\,\,.
 \ee 

\bea \label{K^i_je}
\left(K^i_{\,\,\,j}\right)\,\,= \left(K^{(i)}_{\,\,\,(j)}\right) &=&
\,\,  \left(
\matrix{ -\frac{1}{N}\left(\frac{\partial_t A}{A}
  + \frac{N^r \partial_r A}{A} +  \partial_r N^{r} \right)
   & 0 & 0 \cr 0 & -\frac{1}{N}\left( 
   {\frac{\partial_t B}{B}} + 
   {\frac{N^{r}\,\partial_r B}
     {B}}+ {\frac{N^{r}}{r}} \right) & 0 \cr 0 & 0 &
 -\frac{1}{N}\left( 
   {\frac{\partial_t B}{B}} + 
   {\frac{N^{r}\,\partial_r B}
     {B}} +  {\frac{N^{r}}{r}}\right)  \cr  } \right) \nonumber \\
&=&
\,\,  \left(
\matrix{ -\frac{1}{N}\left(\frac{\partial_t A}{A}
 +   \frac{\partial_r N^{(r)}}{A}\right)
   & 0 & 0 \cr 0 & -\frac{1}{N}\left(  
   {\frac{\partial_t B}{B}} + 
   {\frac{N^{(r)}\,\partial_r B}
     {A\,B}} + {\frac{N^{(r)}}{r\,A}} \right) & 0 \cr 0 & 0 &
 -\frac{1}{N}\left( 
   {\frac{\partial_t B}{B}} + 
   {\frac{N^{(r)}\,\partial_r B}
     {A\,B}} + {\frac{N^{(r)}}{r\,A}} \right)  \cr  } \right) 
\,\,\,\,.
 \eea 
where the index of Eq. (\ref{K^i_je}) was raised with $h^{ij}$ from 
Eq. (\ref{K_ijsph}).

The three-scalar of curvature is given by
\be
^3 R = \frac{2}{r^2A^2}\left(\frac{A^2}{B^2}-1\right) 
+ \frac{2}{A^2}\left( \frac{2\partial_r A}{rA}+ \frac{2(\partial_r B)
(\partial_r A)}{BA} - \frac{(\partial_r B)^2}{B^2} - 
\frac{6\partial_r B}{rB}- \frac{2\partial^2_{rr}B}{B}\right)
\,\,\,\,.
\ee

\subsection{The 3+1 Einstein equations}
\bigskip

It is useful to introduce the new variables,
\bea
\label{nu}
\nu &:=& {\rm ln}[N]\,\,\,\,,\\
\label{alfa}
\alpha &:=& {\rm ln}[A]\,\,\,\,,\\
\label{beta}
\beta &:=& {\rm ln}[B]\,\,\,\,.
\eea

The Hamiltonian constraint Eq.(\ref{CEHf}) reads
\bea
&&
\frac{1}{r^2}\left(\frac{A^2}{B^2} -1\right)  + A^2\left[ 
  2 K^{(r)}_{\,\,\,\,(r)}\,K^{(\theta)}_{\,\,\,\,(\theta)}  + 
  (K^{(\theta)}_{\,\,\,\,(\theta)})^2\right] \nonumber \\
\label{ham3}
&& 
  + \frac{2\,\partial_r \alpha}{r} - 
   \frac{6\,\partial_r \beta}{r} + 
  2(\partial_r \alpha)\,(\partial_r \beta) - 
  3(\partial_r \beta)^2 - 2\,\partial^2_{rr} \beta= 8\pi G_0 E A^2 \,\,\,\,.
\eea

where we have used the fact that the non-diagonal components of 
$K^i_{\,\,\,j}$ are null and that $ K^{(\theta)}_{\,\,\,\,(\theta)}
= K^{(\phi)}_{\,\,\,\,(\phi)}$ [cf. Eq. (\ref{K^i_je})].

The radial component of the momentum constraints Eq.(\ref{CEMf}), 
reads
\be\label{J_r4}
\left(K^{(r)}_{\,\,\,\,(r)} 
  - K^{(\theta)}_{\,\,\,\,(\theta)} \right) 
\left(\frac{1}{r} + \partial_r \beta \right) \nonumber \\
-  \partial_r K^{(\theta)}_{\,\,\,\,(\theta)}
= 4\,\pi G_0 \,A\,J_{(r)}\,\,\,\,,
\ee 
Or in terms of the trace $K$,
\be\label{J_r5}
2\left(K^{(r)}_{\,\,\,\,(r)} 
  - K^{(\theta)}_{\,\,\,\,(\theta)} \right) 
\left(\frac{1}{r} + \partial_r \beta \right) \nonumber \\
-  \partial_r K +  \partial_r K^{(r)}_{\,\,\,\,(r)} 
= 8\,\pi G_0 \,A\,J_{(r)}\,\,\,\,.
\ee 

The angular components of Eq.(\ref{CEMf}) and the spherical symmetry 
lead to the conditions $J_\theta=0=J_\phi$ which implies the absence of 
``angular currents''. 

The dynamical equations for the non-diagonal components of Eq.(\ref{EDEf}), 
i.e., for 
$\partial_t  K^{(r)}_{\,\,\,\,(\theta)}$, 
  $\partial_t  K^{(r)}_{\,\,\,\,(\phi)}$, 
$\partial_t  K^{(\theta)}_{\,\,\,\,(\phi)}$ 
with the fact that $K^{(r)}_{\,\,\,\,(\theta)}= 
K^{(r)}_{\,\,\,\,(\phi)}= K^{(\theta)}_{\,\,\,\,(\phi)}= 0$ 
[cf. Eq. (\ref{K^i_je}) ], leads respectively to the
conditions that $S^{(r)}_{\,\,\,\,(\theta)}= S^{(r)}_{\,\,\,\,(\phi)}=
S^{(\theta)}_{\,\,\,\,(\phi)}= 0$. Moreover, 
taking into account
 the fact that 
$K^{(\theta)}_{\,\,\,\,(\theta)}= K^{(\phi)}_{\,\,\,\,(\phi)}$, 
the dynamical equations for 
$\partial_t K^{(\theta)}_{\,\,\,\,(\theta)}$ and 
$\partial_t K^{(\phi)}_{\,\,\,\,(\phi)}$ [see Eq.(\ref{EDEf})], 
leads to the condition 
$ S^{(\theta)}_{\,\,\,\,(\theta)}=  
S^{(\phi)}_{\,\,\,\,(\phi)}$ corresponding to 
an ``isotropic'' energy-momentum tensor which is 
compatible with the hypothesis of a spacetime with 
spherical symmetry. In this way, the only two non-trivial 
dynamical equations are
\bea
&&
 \partial_t K^{(r)}_{\,\,\,\,(r)} + 
{\frac{N^{(r)}\,\partial_r K^{(r)}_{\,\,\,\,(r)}}{A}} -
N K K^{(r)}_{\,\,\,\,(r)}  - \frac{N}{A^2}\left(
 \frac{2\partial_r \alpha}{r} - 
 \frac{4\partial_r \beta}{r} + 
 2(\partial_r \alpha)\,(\partial_r \beta)
 - 2(\partial_r\beta)^2
- 2\partial^2_{rr} \beta \right) \nonumber \\
\label{dynk113}
&& + \frac{N}{A^2}\left[\partial^2_{rr}\nu + (\partial_r \nu)^2
-(\partial_r \alpha)\,(\partial_r \nu)  \right]    
 =  4\,\pi G_0 \,N\,\left( -S^{(r)}_{\,\,\,\,(r)} + 
  2 S^{(\theta)}_{\,\,\,\,(\theta)} - E\right) \,\,\,\,,
\eea

\bea
&&  \partial_t K^{(\theta)}_{\,\,\,\,(\theta)} + 
\frac{N^{(r)}\,\partial_r K^{(\theta)}_{\,\,\,\,(\theta)}}{A} 
-NK K^{(\theta)}_{\,\,\,\,(\theta)}
 - \frac{N}{A^2}\left[ \frac{1}{r^2}\left(\frac{A^2}{B^2}-1\right)
 + \frac{\partial_r \alpha}{r}  
  - \frac{4\partial_r \beta}{r}  
  + (\partial_r \alpha)\,(\partial_r \beta)
  - 2(\partial_r \beta)^2 
 - \partial^2_{rr} \beta  
 \right] \nonumber \\ 
\label{dynk223}
&& 
+ \frac{N}{A^2}\left[ (\partial_r \beta)\,(\partial_r \nu)
+ \frac{\partial_r \nu}{r}\right]
=   4\,\pi G_0 \,N\,\left(S^{(r)}_{\,\,\,\,(r)}  - E\right) \,\,\,\,.
\eea

From Eq. (\ref{EDK2}) the evolution equation for the trace of $K_{ij}$ is

\bea
&&  \partial_t K + \frac{N^{(r)}\,\partial_r K}{A}
- N \left[ (K^{(r)}_{\,\,\,\,(r)})^2 + 
2 (K^{(\theta)}_{\,\,\,\,(\theta)})^2  
 \right] \nonumber \\ 
\label{dynK4}
&& 
+\frac{N}{A^2}\left[ \partial^2_{rr}\nu + (\partial_r\nu)^2
- (\partial_r \alpha)\,(\partial_r \nu) 
+2(\partial_r \beta)\,(\partial_r \nu)
+ \frac{2\partial_r \nu}{r}\right]
=   4\,\pi G_0 \,N\,\left(S + E\right) \,\,\,\,.
\eea

where we recognize the 3-covariant Laplace operator in  
spherical coordinates of the slices $\Sigma_t$ [see Eq. (\ref{ds})]:
\be
\,^3\Delta \nu = \frac{1}{A^2}\left[
 \partial^2_{rr}\nu + \frac{2\partial_r \nu}{r} 
- (\partial_r \alpha)\,(\partial_r \nu)
+2(\partial_r \beta)\,(\partial_r \nu)\right] \,\,\,.
\ee

\subsection{Matter equations}

In the present case of a perfect fluid in spherical symmetry we have
\bea \label{Er}
 E_{\rm PF} &=& (\rho +p)\Gamma -p \,\,\,\,,\\
\label{Srr}
_{\rm PF}\,\!S^{(r)}_{\,\,\,(r)} &=& (E_{\rm PF} + p ) (\,^3\,\!U^{(r)})^2 + p \,\,\,\,,\\
\label{Sthth}
_{\rm PF}\,\!S^{(\theta)}_{\,\,\,(\theta)} &=& S^{(\phi)}_{\,\,\,(\phi)} = p\,\,\,\,\\
\label{JPPFr}
J^{\rm PF}_{(r)} &=& (E_{\rm PF} + p )\,\,^3\,\!U^{(r)} \,\,\,\,,\\
\label{lorentzr}
\Gamma &=& \left[ 1- (\,^3\,\!U^{(r)})^2\right]^{-1/2}\,\,\,\,,
\eea
where 
\be \label{P3Ur}
\,^3\,\!U^{(r)} = 
\frac{A}{N}\left(V^r- N^r\right) 
= \frac{1}{N}\left(V^{(r)}- N^{(r)}\right) \,\,\,\,,
\ee
with $V^r:= u^r/u^t$. The spherical symmetry implies $u^\theta=0=u^\phi$ and 
therefore $U^{(\theta)}=0= U^{(\phi)}$.

Note that 
\be\label{SrrJ}
_{\rm PF}\,\!S^{(r)}_{\,\,\,(r)}= \,^3\,\!U^{(r)} J^{\rm PF}_{(r)} + p\,\,\,.
\ee

The Eq. (\ref{nrjEuler}) reads
\be\label{ECEfss}
\partial_t E_{\rm PF} +  N^r\partial_r E_{\rm PF} + \frac{N}{AB^2 r^2}\partial_r
\left(AB^2 r^2 J^r_{\rm PF} \right) = 
N \left( \,_{\rm PF}\,\!S^{(r)}_{\,\,\,\,(r)} K^{(r)}_{\,\,\,\,(r)}
+ 2 \,_{\rm PF}\,\!S^{(\theta)}_{\,\,\,\,(\theta)} K^{(\theta)}_{\,\,\,\,(\theta)}  
+  E_{\rm PF} K\right) - 2 J^r_{\rm PF} \partial_r N - N^2  {\cal F}^t
\,\,\,\,.
\ee
Or in terms of $J_{(r)}= A J^r$, 
\be\label{ECEfss2}
\partial_t E_{\rm PF} +  N^r\partial_r E_{\rm PF} + \frac{N}{AB^2 r^2}\partial_r
\left(B^2 r^2 \,J^{\rm PF}_{(r)}\right) = 
N \left( \,_{\rm PF}\,\!S^{(r)}_{\,\,\,\,(r)} K^{(r)}_{\,\,\,\,(r)}  
+ 2 \,_{\rm PF}\,\!S^{(\theta)}_{\,\,\,\,(\theta)} K^{(\theta)}_{\,\,\,\,(\theta)} 
+  E_{\rm PF}K\right) - \frac{2}{A} \,J^{\rm PF}_{(r)} \partial_r N - N^2  {\cal F}^t
\,\,\,\,.
\ee
Using Eqs. (\ref{Sthth}), (\ref{P3Ur}) and (\ref{SrrJ}) 
the latter equation can be written in conservative form as
\bea
\partial_t E_{\rm PF} + \frac{1}{r^2}\partial_r
\left(r^2 V^r \,E_{\rm PF}\right) &=& E_{\rm PF}\left( N K + 
\frac{2N^r}{r} + \partial_r N^r\right) - 
\frac{1}{r^2}\partial_r
\left( r^2 \frac{N}{A} \,^3\,\!U^{(r)}\,p\right) \nonumber \\
\label{ECEcons}
&& -\frac{N J^{\rm PF}_{(r)}}{A}\left[ \partial_r\nu + \partial_r\alpha + 
2\partial_r\beta - \,^3\,\!U^{(r)}\,A K^{(r)}_{\,\,\,\,(r)} \right] 
+ N K p  - N^2 {\cal F}^t
\,\,\,\,.
\eea

The momentum conservation Eq.(\ref{ECMf}) reads
\be\label{ECMfss}
\partial_t \,J^{\rm PF}_r + N^r\partial_r \,J^{\rm PF}_r 
+ \,J^{\rm PF}_r\partial_r N^r + N (\,^3\nabla_l \,_{\rm PF}\,\!S^l_{\,\,\,r})
=   NK \,J^{\rm PF}_r -
\left( \,_{\rm PF}\,\!S_{\,\,\,r}^r + E_{\rm PF} \right)\,\partial_r N   
- \, ^3{\cal F}_r N \,\,\,\,.
\ee

Explicitly
\bea \label{ECMfss2}
&&
 \partial_t \,J^{\rm PF}_{r}  +  N^r\partial_r \,J^{\rm PF}_r + 
\,J^{\rm PF}_r\partial_r N^r 
 + N\,\partial_r \,_{\rm PF}\,\!S^{(r)}_{\,\,\,\,(r)}
 + 2N\left( \,_{\rm PF}\,\!S^{(r)}_{\,\,\,\,(r)} - 
\,_{\rm PF}\,\!S^{(\theta)}_{\,\,\,\,(\theta)}\right)\left( \frac{1}{r} + 
 \frac{\partial_r B }{B} \right) 
\nonumber \\
&&  = -\left( \,_{\rm PF}\,\!S^{(r)}_{\,\,\,\,(r)} + 
  E_{\rm PF} \right) \partial_r N  
+ N\, \,J^{\rm PF}_{r}\left( K^{(r)}_{\,\,\,\,(r)} 
       + 2\,\, K^{(\theta)}_{\,\,\,\,(\theta)} \right) 
- \, ^3{\cal F}_r N \,\,\,.
\eea
Or in terms of $J_{(r)}$, we have
\bea \label{ECMfss3}
&&
 \partial_t \,J^{\rm PF}_{(r)}  + \,J^{\rm PF}_{(r)}\frac{\partial_t A}{A} + 
\frac{N^{(r)}}{A}\,\partial_r \,J^{\rm PF}_{(r)}  + 
\frac{\,J^{\rm PF}_{(r)}}{A} \partial_r N^{(r)}
 + \frac{N}{A}\,\partial_r \,_{\rm PF}\,\!S^{(r)}_{\,\,\,\,(r)}
 + \frac{2N}{A}\left(\,_{\rm PF}\,\!S^{(r)}_{\,\,\,\,(r)} - 
\,_{\rm PF}\,\!S^{(\theta)}_{\,\,\,\,(\theta)}\right)\left( \frac{1}{r} + 
 \frac{\partial_r B }{B} \right) 
\nonumber \\
&&  = -\left(\,_{\rm PF}\,\!S^{(r)}_{\,\,\,\,(r)} + 
  E_{\rm PF} \right) \frac{\partial_r N}{A}  + 
N\,\,J^{\rm PF}_{(r)}\left( K^{(r)}_{\,\,\,\,(r)} 
       + 2\,\, K^{(\theta)}_{\,\,\,\,(\theta)} \right) 
-\, ^3{\cal F}_{(r)} N \,\,\,.
\eea
When using Eq. (\ref{K^i_je}) to replace the time derivative of $A$ we find
\bea \label{ECMfss4}
&&
 \partial_t \,J^{\rm PF}_{(r)}  + 
\frac{N^{(r)}}{A}\,\partial_r \,J^{\rm PF}_{(r)}  
 + \frac{N}{A}\,\partial_r \,_{\rm PF}\,\!S^{(r)}_{\,\,\,\,(r)}
 + \frac{2N}{A}\left(\,_{\rm PF}\,\!S^{(r)}_{\,\,\,\,(r)} - 
\,_{\rm PF}\,\!S^{(\theta)}_{\,\,\,\,(\theta)}\right)\left( \frac{1}{r} + 
 \frac{\partial_r B }{B} \right) 
\nonumber \\
&&  = -\left(\,_{\rm PF}\,\!S^{(r)}_{\,\,\,\,(r)} + 
  E_{\rm PF} \right) \frac{\partial_r N}{A}  + 
2N\,\,J^{\rm PF}_{(r)}\left( K^{(r)}_{\,\,\,\,(r)} 
       + \,\, K^{(\theta)}_{\,\,\,\,(\theta)} \right)
- \, ^3{\cal F}_{(r)} N \,\,\,.
\eea

Using Eqs. (\ref{P3Ur}) and (\ref{SrrJ}), one obtains an equation for 
$J^{\rm PF}_{(r)}$ in conservative form 
\bea\label{Jcons}
 \partial_t \,J^{\rm PF}_{(r)}  + \frac{1}{r^2}\partial_r\left(
r^2 V^r J^{\rm PF}_{(r)}\right) &=& \frac{N}{A}
\left\{J^{\rm PF}_{(r)}\left[ 2A\left(
 K^{(r)}_{\,\,\,\,(r)} 
       + \,\, K^{(\theta)}_{\,\,\,\,(\theta)} \right)
 -\,^3\,\!U^{(r)}\left(\partial_r\alpha + 2 \partial_r\beta\right)
 -\frac{A}{N}\left(\frac{2}{r}N^r + \partial_r N^r\right)\right]\right.
\nonumber \\
&& \left. - \left(E_{\rm PF}+ p\right)\partial_r\nu 
-\partial_r p - \, ^3{\cal F}_{(r)} A\right\} \,\,\,.
\eea
with the alternative form for the r.h.s,
\bea\label{JconsV}
 \partial_t \,J^{\rm PF}_{(r)}  + \frac{1}{r^2}\partial_r\left(
r^2 V^r J^{\rm PF}_{(r)}\right) &=& 
J^{\rm PF}_{(r)}\left[ 2N\left(
 K^{(r)}_{\,\,\,\,(r)} 
       + \,\, K^{(\theta)}_{\,\,\,\,(\theta)} \right)
 - V^r\left(\partial_r\alpha + 2 \partial_r\beta\right)
 + N^r\left(\partial_r\alpha + 2 \partial_r\beta -\frac{2}{r}\right)
 -  \partial_r N^r\right]
\nonumber \\
&&  - \frac{N}{A}\left[\left(E_{\rm PF}+ p\right)\partial_r\nu 
+\partial_r p + \, ^3{\cal F}_{(r)} A \right]  \,\,\,.
\eea
Another possibility which has turned to be very useful in some numerical studies 
\cite{gour1,gour2,gour3}, is the use of the Euler equation for the fluid instead 
of the equation for $J^{\rm PF}_{(r)}$. 
The only non-trivial component of the Euler equation (\ref{EulerRG}) 
in spherical symmetry reads
\bea \label{EulerRGr}
& & \partial_{(t)} \,^3\,\!U^{(r)} + \,^3\,\!U^{(r)}\,\,
^3 \partial_{(r)}\,^3\,\!U^{(r)}  = 
- \frac{1}{E_{\rm PF}+p} \left[ \,^3\partial^{(r)}p + \,^3\,\!U^{(r)} \partial_{(t)}p 
\right]  \nonumber \\
\!\!\!\!\!\!\!\!\!\!\!\!
&+& \frac{1}{\Gamma^2}\left(\,\,^3\,\!U^{(r)} K^{(r)}_{\,\,\,(r)}- a_{(r)}\,\right) 
+ \frac{1}{E_{\rm PF}+p}\left(\,^3\,\!U^{(r)}{\cal F}^{(t)} - {\cal F}^{(r)}\right)
\,\,\,\,.
\eea
where we used the fact that the RRC are antisymmetric so that 
the terms ${\cal O}^{(r)}_{(t)(r)}=0= {\cal O}^{(r)}_{(r)(r)}$ and we remind the 
explicit expressions
\bea
\partial_{(t)} &=& \frac{1}{N}\partial_t + \frac{N^r}{N}\partial_r \,\,\,,\\
\,^3\,\!\partial_{(r)} &=& \frac{1}{A}\partial_r\,\,\,\,,\\
a_{(r)} &=& \,^3\,\!\partial_{(r)} \nu \,\,\,.
\eea

\bigskip
\subsection{The RBE in spherical symmetry}
\bigskip

The most general 4-metric for a spherically symmetric spacetime is given 
by Eq.(\ref{ds}). Then, it's easy to see that the simplest tetrad 
choice $e_{(\mu)}$ associated to Eq.(\ref{ds}) and which 
corresponds to the local tetrad of the Eulerian observer, reads,
\bea
 e_{(t)} &=& \frac{1}{N}\frac{\partial}{\partial t} + \frac{N^r}{N}
\frac{\partial}{\partial r}\,,\,\,\\
 e_{(r)} &=& \frac{1}{A}\frac{\partial}{\partial r} \,,\,\, \\  
 e_{(\theta)} &=& \frac{1}{rB}\frac{\partial}{\partial \theta} \,,\,\,\\   
 e_{(\phi)} &=& \frac{1}{rB\sin\theta}\frac{\partial}{\partial \phi} 
\,\,\,\,.  
\eea
where $\partial/\partial x^\mu$ [with $x^\mu=(t,r,\theta,\phi)$] denote the 
coordinate basis.

In the Eulerain frame, the spherical symmetry of configuration space (i.e., spacetime) 
induces a symmetry on momentum space that can be exploited to simplify the computations.
It is convenient to define spherical variables on momentum space as 
follows: take a unit vector $e_{(r)}$ 
as polar axis, i.e., as symmetry axis on momentum space. 
Then it is useful to introduce new variables  $e, \psi, \gamma$ in the Eulerian frame as 
\be \label{impspher}
p^{(t)}= e\,,\,\,p^{(r)}= p \cos\psi\,,\,\, p^{(\theta)}= 
p\sin\psi \cos \gamma\,,\,\,
p^{(\phi)}=p\sin\psi \sin \gamma \,\,\,\,.
\ee
As mentioned earlier in Sec. IV, $e$ is the energy of the radiated particles (e.g., 
neutrinos) in the Eulerian frame and $p^2= p^{(i)} p_{(i)}$; 
$\psi$\, is the angle between the polar axis and the 
neutrino propagation three-vector $p^{(i)}$, 
and \,$\gamma$\, is the angle of rotation 
around $e_{(r)}$. The above variables are consitent with the 
mass shell condition (\ref{curvms2}). For massless 
particles, then $p\equiv e$. Under these new variables the RBE (\ref{Boltzeq4}) reads
\be \label{Boltzeq5}
\left(\,p^{(a)}(e,\psi,\gamma)q^{\mu}_{(a)}\frac{\partial}{\partial x^{\mu}} - 
p^{(b)}(e,\psi,\gamma)p^{(a)}(e,\psi,\gamma){\cal O}^{(d)}_{(b)(a)}
\frac{\partial \tilde p^{(c)}}{\partial p^{(d)}}
\frac{\partial }{\partial \tilde p^{(c)}}
\,\,\right)F\left(
\tilde{\mbox{\bm{$p$}}}\right)
 = \left(\frac{d F}
{d \tau}\right)_{\rm coll} \,\,\,\,,
\ee
\noindent{where} we have explicitly stress the dependence of 
momenta with respect to the new momenta spherical-like 
variables $(e,p,\psi,\gamma)$ represented colectively by
 $\tilde{\mbox{\bm{$p$}}}$. 

Indeed, the spherical symmetry and the mass shell condition 
will be reflected in the RBE by the fact that 
the distribution function \,$F_{\rm R}$ depends only of 
four phase-space coordinates ($t,r,e,\mu$) or ($t,r,p,\mu$) 
instead of the original eight ($x^\mu,p^\mu$) \cite{Lind,Schinder}, where $\mu:= \cos\psi$ . 

\vskip .3cm
We can now proceed to calculate explicitly the RBE.
The only non-null Ricci coefficients are \cite{3+1math},
\bea
{\cal O}^{(t)}_{(t)(r)} &=& {\cal O}^{(r)}_{(t)(t)}=  D_r\nu \,\,,\,\,
{\cal O}^{(t)}_{(r)(r)}= {\cal O}^{(r)}_{(r)(t)}= D_t\alpha + \frac{1}{N}D_r N^{(r)} \,\,,\,\,\nonumber \\
{\cal O}^{(t)}_{(\theta)(\theta)} &=& {\cal O}^{(t)}_{(\phi)(\phi)}= 
{\cal O}^{(\theta)}_{(\theta)(t)} = {\cal O}^{(\phi)}_{(\phi)(t)}=
\frac{N^{(r)}}{r A N} + D_t\beta + \frac{N^{(r)}} {N}D_r\beta \,\,,\,\,
\nonumber \\
{\cal O}^{(r)}_{(\theta)(\theta)} &=&
{\cal O}^{(r)}_{(\phi)(\phi)}= -{\cal O}^{(\theta)}_{(\theta)(r)}= -{\cal O}^{(\phi)}_{(\phi)(r)}=
-r^{-1}A^{-1} - D_r\beta \,\,,\,\,\nonumber \\
{\cal O}^{(\theta)}_{(\phi)(\phi)} &=& -{\cal O}^{(\phi)}_{(\phi)(\theta)}= 
-\frac{1}{r B}\cot\theta\,,\,\,\nonumber 
\eea

\noindent{where} $
D_t:= \frac{1}{N}\frac{\partial}{\partial t}\,,\,\,
D_r:= \frac{1}{A}\frac{\partial}{\partial r} \equiv \,^3\partial_{(r)}$. 
After imposing the mass shell condition, 
the RBE (\ref{Boltzeq5}) in terms of the spherical variables reads 
explicilty \cite{3+1math},
\bea
& &
 e D_t F_{\rm R}(r,t,e,\mu) +\left(\frac{p\,\mu\,}{e}  
+ \frac{N^{(r)}}{N}\right)e D_r F_{\rm R}(r,t,e,\mu) \nonumber \\
& & 
-p^2 \left[(1-\mu^2){\frac{N^{(r)}\,}{N}}\left(\frac{1}{rA}+D_r\beta\right) 
+ \mu^2 D_t\alpha 
+ (1-\mu^2)D_t\beta
+ \frac{\mu e}{p} D_r\nu
+ {\frac{{\mu}^2}{N}}D_r N^{(r)}\right]\partial_e F_{\rm R}(r,t,e,\mu)\nonumber \\
&& 
+ e(1-\mu^2)
\left[\left({\frac{p\,}{e}} + {\frac{\mu\,N^{(r)}\,}{N}}\right)
\left(\frac{1}{rA} + D_r\beta\right) +\mu\left(
 D_t\beta - D_t\alpha\right) - \frac{e}{p} D_r\nu
-{\frac{\mu\,}{N}} D_r N^{(r)} \right]\partial_\mu F_{\rm R}(r,t,e,\mu) \nonumber \\
& &
= \left( \frac{d F}{d\lambda}\right)_{\rm coll} \,\,\,\,.
 \eea

The alternative 3+1 form of the RBE can be computed from (\ref{3+1RBE1}) 
when changing to the spherical variables $\tilde p^{(\mu)}$ 
in the momentum space:
\bea
\label{3+1RBEss}
& &
 e D_t F_{\rm R}(r,t,E,\mu) +\left(\frac{p\,\mu\,}{e}  
+ \frac{N^{(r)}}{N}\right)e D_r F_{\rm R}(r,t,e,\mu) 
-p^2 \left[-(1-\mu^2)K^{(\theta)}_{\,\,\,\,\,(\theta)} -\mu^2 K^{(r)}_{\,\,\,\,\,(r)}  
+ \frac{\mu e}{p} D_r\nu \right]\partial_e F_{\rm R}(r,t,e,\mu)\nonumber \\
&& 
+ e(1-\mu^2)
\left[{\frac{p\,}{e}} 
\left(\frac{1}{rA} + D_r\beta\right) +\mu\left(K^{(r)}_{\,\,\,\,\,(r)} - 
K^{(\theta)}_{\,\,\,\,\,(\theta)}\right) -\frac{e}{p}D_r\nu
 \right]\partial_\mu F_{\rm R}(r,t,e,\mu) 
= \left( \frac{d F}{d\lambda}\right)_{\rm coll} \,\,\,\,.
 \eea

In normal coordinates where $N^r= 0$, and for massless particles, 
Eq.(\ref{Boltzeq5}) reduces to 
the relativistic Boltzman equation derived by Lindquist\cite{Lind} 
with the choice 
$B=R(r)/r$. Under the isotropic gauge Eq.(\ref{3+1RBEss}) can be 
written in a ``conservative'' form which is specially suited for 
numerical solutions \cite{HH}. Following Harleston \& Vishniac 
\cite{HV}, we can write Eq.(\ref{3+1RBEss}) in conservative form 
as
\be\label{Boltzcons}
\partial_t \tilde F + \frac{1}{r^2}\partial_r  
\left(r^2 V_p^r \tilde F\right) + \frac{1}{pe}\partial_e 
\left(p^3 {\cal H}_e \tilde F\right) + \partial_\mu
\left({\cal H}_\mu \tilde F\right) = \frac{NAB^2}{e} 
\left( \frac{d F}{d\lambda}\right)_{\rm coll}
\ee
where
\bea
\label{Fconf}
\tilde F &=& AB^2 F_{\rm R}(r,t,e,\mu) \,\,\,\,\,,\\
 V_p^r &=& \frac{p^r}{p^t}= \frac{N}{A}\left(\frac{p\mu}{e} +
\frac{N^{(r)}}{N}\right)\,\,\,,\\
{\cal H}_e &=& \frac{N}{p^2}\hat{\mbox{\bm{$L$}}} (e) 
= -N \left[-(1-\mu^2)K^{(\theta)}_{\,\,\,\,\,(\theta)} -\mu^2 K^{(r)}_{\,\,\,\,\,(r)}  
+ \frac{\mu e}{p} D_r\nu \right]
\,\,\,\,\\
\label{Hmu}
{\cal H}_\mu &=& \frac{N}{e}\hat{\mbox{\bm{$L$}}}(\mu)
= N (1-\mu^2)
\left[{\frac{p\,}{e}} 
\left(\frac{1}{rA} + D_r\beta\right) +\mu\left(K^{(r)}_{\,\,\,\,\,(r)} - 
K^{(\theta)}_{\,\,\,\,\,(\theta)}\right) -\frac{e}{p}D_r\nu
 \right]
\,\,\,\,,
\eea
and $\hat{\mbox{\bm{$L$}}}$ stands for the Liouville operator 
as it appears in the l.h.s of Eq.(\ref{3+1RBEss}). Alternatively, 
the above equation can be written as
\be\label{Boltzcons2}
\partial_t \tilde F + \frac{1}{r^2}\partial_r  
\left(r^2 V_p^r \tilde F\right) + \frac{1}{p^2}\partial_p 
\left(p^3 {\cal H}_p \tilde F\right) + \partial_\mu
\left({\cal H}_\mu \tilde F\right) = \frac{NAB^2}{e} 
\left( \frac{d F}{d\lambda}\right)_{\rm coll}\,\,\,,
\ee
where now $\tilde F= \tilde F_{\rm R}(r,t,p,\mu) $ is to be regarded as a function of 
$p$ instead of $e$ and 
\be
\label{Hp}
{\cal H}_p = \frac{N}{pe}\hat{\mbox{\bm{$L$}}}(p) 
= -N \left[-(1-\mu^2)K^{(\theta)}_{\,\,\,\,\,(\theta)} -\mu^2 K^{(r)}_{\,\,\,\,\,(r)}  
+ \frac{\mu e}{p} D_r\nu \right]\,\,\,\,.
\ee

When one of the different gauges and slicing conditions are chosen 
to write the RBE this will only affect the particular form of the quantities 
given by (\ref{Fconf})$-$(\ref{Hmu}) and (\ref{Hp}), and the r.h.s of 
Eqs.(\ref{Boltzcons}) and (\ref{Boltzcons2}).

Finally, we emphasize that all the momentum variables which appear in the 
various forms of the RBE in spherical symmetry, are components with respect to the 
orthonormal tetrad carried by the Eulerian observer. 
Therefore, the corresponding quantities in the 
collision integral are to be referred to the same observer.
\vskip .3cm

\bigskip
\subsection{Mean radiative variables}
\bigskip
In Sec. IV we defined the energy-momentum tensor of particles 
in terms of their microscopic four-momenta. However, we can introduce 
mean radiative variables which are to be interpreted as their counterparts  
of the continuum case. Such variables are called the 
{\it moments of the distribution function}. 
In order to write them explicitly, we shall use the physical components of 
the energy-momentum tensor of particles Eq. (\ref{Tparttet}) 
and the spherical variables in momentum space in addition to the 
invariant volume element of momentum space given by Eq.(\ref{dPsph}). 
We have then
\bea 
T^{(\mu)(\nu)}_{\rm R} &=&  
 \int^\infty _{0} \int^1_{-1} \int^{2\pi}_0 p^{(\mu)}p^{(\nu)} F_{\rm R}(r,t,p,\mu)
\frac{p^2}{e} dp d\mu d\gamma \nonumber \\
\label{EMNtensor}
&=& \int^\infty _{\tilde m} \int^1_{-1} \int^{2\pi}_0 p^{(\mu)}p^{(\nu)} F_{\rm R}(r,t,e,\mu)
\sqrt{e^2- \tilde m^2} de d\mu d\gamma 
\,\,\,\,.
\eea
The only non-null moments are
\bea \label{moment1}
E_{\rm R} &:=& T^{(t)(t)}_{\rm R} = 
2\pi \int^\infty _{0}  p^2 \sqrt{p^2+\tilde m^2}  dp \int^1_{-1}   F_{\rm R}(r,t,p,\mu) d\mu
= 2\pi \int^\infty _{\tilde m} e^2 \sqrt{e^2-\tilde m^2} de \int^1_{-1} 
F_{\rm R}(r,t,e,\mu) d\mu \,\,\,\,,\\
\label{Hflux}
H_{\rm R} &:=& T^{(t)(r)}_{\rm R} = 2\pi \int^\infty _{0} p^3 dp \int^1_{-1} \mu F_{\rm R}(r,t,e,\mu) d\mu
= 2\pi \int^\infty _{\tilde m} e (e^2-\tilde m^2) de 
\int^1_{-1} \mu F_{\rm R}(r,t,e,\mu) 
d\mu 
\,\,\,\,,\\
p_{\rm R} &:=& T^{(r)(r)}_{\rm R} = 2\pi \int^\infty _{0} \frac{p^4}{\sqrt{p^2 +\tilde m^2}}dp
\int^1_{-1} \mu^2 F_{\rm R}(r,t,e,\mu) d\mu
= 2\pi \int^\infty _{\tilde m}  \left(e^2-\tilde m^2\right)^{3/2} de \int^1_{-1} 
\mu^2 F_{\rm R}(r,t,e,\mu) d\mu
\,\,\,\,.
\eea
The tangential preassures are given by,
\bea
p^T_{\rm R}:= T^{(\theta)(\theta)}_{\rm R} =T^{(\phi)(\phi)}_{\rm R} &=& 
\pi \int^\infty _{0} \frac{p^4}{\sqrt{p^2 +\tilde m^2}}dp
\int^1_{-1} (1- \mu^2) F_{\rm R}(r,t,e,\mu) d\mu \nonumber \\
 &=&
 \pi \int^\infty _{\tilde m}  \left(e^2-\tilde m^2\right)^{3/2} de \int^1_{-1} 
(1-\mu^2) F_{\rm R}(r,t,e,\mu) d\mu
\eea

We note that in the massless case,
\be
p^T_{\rm R} = \frac{1}{2} (E_{\rm R} - p_{\rm R}) \,\,\,\,.
\ee

The corresponding 3+1 variables are
\bea
J^{(r)}_{\rm R} &=& H_{\rm R}\,\,\,\,,\\
S^{(r)(r)}_{\rm R} &=& p_{\rm R} \,\,\,\,,\\
S^{(\theta)(\theta)}_{\rm R} &=& S^{(\phi)(\phi)}_{\rm R}= p^T_{\rm R} \,\,\,,\\
S_{\rm R} &=& \,_{\rm R}\,\!S^{(i)}_{\,\,\,\,(i)}= p_{\rm R} + 2 p^T_{\rm R}   \,\,\,\,.
\eea
The effective pressure of radiation which can be defined as 
 $p_{\rm eff}^{\rm R}= S_{\rm R}/3$ turns to be in the case of massless particles 
$p_{\rm eff}^{\rm R}= E_{\rm R}/3$ which corresponds 
precisely to the equation of state (EOS) of an ultrarelativistic gas. In 
Eq. (\ref{Hflux}),  
$H_{\rm R}$ is the mean radiative flux of energy in the radial direction. 

It is usual to introduce the so called variable Eddington factor 
\be
\Xi = \frac{ p_{\rm R}}{E_{\rm R}}\,\,\,,
\ee
used to measure the degree of ``anisotropy'' in the particle flow. In the 
case of massless particles if $\Xi=1/3$ then  $p^T_{\rm R}= p_{\rm R}=E_{\rm R}/3$ 
which corresponds to an fully isotropic flow. 
Moreover, in the free streaming approximation we have
\be\label{free}
E_{\rm R}= J^{(r)}_{\rm R}= S^{(r)(r)}_{\rm R}\,\,\,,
\ee
and so $\Xi=1$, which is the case of a highly anisotropic flow ($p^T_{\rm R}=0$) with a 
purely radial flux of radiation.

In the same way, the macroscopic particle number density current 
measured in the Eulerian frame is given by 
Eq. (\ref{partfluxtet}) and it in terms of the spherical variables of 
momentum space gives
\bea 
j^{(\mu)}_{\rm R} &=&  
 \int^\infty _{0} \int^1_{-1} \int^{2\pi}_0 p^{(\mu)} F_{\rm R}(r,t,p,\mu)
\frac{p^2}{e} dp d\mu d\gamma \nonumber \\
\label{partfluxspher}
&=& \int^\infty _{\tilde m} \int^1_{-1} \int^{2\pi}_0 p^{(\mu)} F_{\rm R}(r,t,e,\mu)
\sqrt{e^2- \tilde m^2} de d\mu d\gamma 
\,\,\,\,.
\eea
In particular, the mean number density and flux of particles measured by the Eulerian observer 
are given respectively by 
\bea 
n_{\rm E}^{\rm R} &:=& -n_\mu j^\mu_{\rm R} = j^{(t)}_{\rm R} =  
2\pi \int^\infty _{0} \int^1_{-1}   F_{\rm R}(r,t,p,\mu)
p^2 dp d\mu \nonumber \\
\label{partfluxspherE}
&=& 2\pi \int^\infty _{\tilde m} \int^1_{-1} \int^{2\pi}_0 e F_{\rm R}(r,t,e,\mu)
\sqrt{e^2- \tilde m^2} de d\mu 
\,\,\,\,,\\
\label{jrRphys}
j^{(r)}_{\rm R} &=& 
2\pi \int^\infty _{0} \frac{p^3}{\sqrt{p^2+ \tilde m^2}} 
dp \int^1_{-1} \mu F_{\rm R}(r,t,e,\mu) d\mu
= 2\pi \int^\infty _{\tilde m} (e^2-\tilde m^2) de 
\int^1_{-1} \mu F_{\rm R}(r,t,e,\mu) 
d\mu \,\,\,\,,
\eea
and $j^{(\theta)}_{\rm R}=0= j^{(\phi)}_{\rm R}$.

In terms of the above macroscopic variables, the energy-momentum conservation 
equation in spherical symmetry Eq. (\ref{conseqs3}) reads according to 
Eqs. (\ref{ECEftet}) and (\ref{ECMftet}) as follows \cite{3+1math}:

\be \label{ECEftetrad}
 \partial_{(t)}E_{\rm R} + \,^3\partial_{(r)} H_{\rm R} - E_{\rm R} K +  2H_{\rm R} 
\left[ \,^3\partial_{(r)}\nu + \frac{1}{rA} + \,^3\partial_{(r)}\beta\right]  - p_{\rm R} K_{(r)(r)} = -{\cal F}^{(t)} \,\,\,\,.
\ee

\be \label{ECMftetrad}
\partial_{(t)} H_{\rm R} + \,^3\partial_{(r)}p_{\rm R} +\left[\frac{1}{rA} + 
\partial_{(r)}\beta\right]\left[3p_{\rm R}- E_{\rm R}\right]
+ \left[ p_{\rm R} + E_{\rm R} \right] \,^3\partial_{(r)}\nu - H_{\rm R}\left( K 
+ K^{(r)}_{\,\,\,(r)}\right) = -{\cal F}^{(i)} \,\,\,\,.
\ee

Depeding on the gauge and slicing choice some of the terms within these 
equations can vanish. 
Actually such evolution equations can be obtained directly from the RBE 
when multiplying this by the momenta and then integrating in momentum 
space.

We emphasize that it is more convenient to calculate 
 $E_{\rm R},H_{\rm R}$ and $p_{\rm R}$, directly from their definition 
once the distribution function has been computed, rather than using the above 
equations. In any case,  
such conservation equations can be used to verify 
the self-consistence of the system. The disadvantage of using the 
system of Eqs. (\ref{ECEftetrad}) and (\ref{ECMftetrad}) for 
the moments of the distribution instead of solving the RBE 
is that such a system is undetermined (i.e, there are more variables 
than equations). Then a closure relation is needed to remove the ambiguity 
(e.g. the diffusion approximation relating $E_{\rm R}$ and $p_{\rm R}$). 
\bigskip

Finally, in spherical symmetry the evolution equations 
(\ref{barcons1}), (\ref{barcons2}), (\ref{entconscoor}), (\ref{entcons3}), 
(\ref{partconscoor}), (\ref{partcons3}) write respectively

\be\label{barcons1r}
\partial_t \left(AB^2 n_E\right) + \frac{1}{r^2}\partial_r\left(r^2 
 V^r n_E AB^2\right) =0\,\,\,,
\ee

\be\label{barcons3}
\partial_t \left( AB^2 n_E\right) + \frac{1}{r^2}\partial_r\left[AB^2 r^2 
n_E \left( N^r + \frac{N}{A}\,^3\,\!U^{(r)}\right)\right] =0\,\,\,,
\ee

\be\label{entconscoorr}
\partial_t \sigma + V^r \partial_r \sigma= 
- \frac{N}{n\Theta\Gamma}\left( \mu^{\rm R} {\cal R}_{\rm R} + {\cal D}_p\right)\,\,\,,
\ee 

\be\label{entcons4}
\partial_t \sigma + N^r\partial_r \sigma + \frac{N \,^3\,\!U^{(r)}}{A} 
\partial_r \sigma=  -\frac{N}{n\Gamma \Theta}\left( \mu^{\rm R} {\cal R}_{\rm R} 
+ {\cal D}_p\right)\,\,\,,
\ee

\be\label{partconscoorr}
\partial_t  x_{\rm R} + V^r \partial_r  x_{\rm R} = \frac{N}{n\Gamma}{\cal R}_{\rm R}\,\,\,,
\ee

\be\label{partcons4}
\partial_t x_{\rm R} + N^r\partial_r x_{\rm R} + \frac{N \,^3\,\!U^{(r)}}{A} 
\partial_r x_{\rm R}
= \frac{N}{n\Gamma}{\cal R}_{\rm R}\,\,\,.
\ee

Equation (\ref{barcons1r}) has a conservative form for the quantity 
$AB^2 n_E$. One can alternatively write an equation in conservative form 
for $n_E$ as 
 \be\label{barcons1r2}
\partial_t n_E + \frac{1}{r^2}\partial_r\left(r^2 
 V^r n_E \right) + n_E\left[\partial_t \alpha + 2\partial_t \beta 
+ V^r \left(\partial_r \alpha + 2\partial_r \beta\right)\right] =0\,\,\,.
\ee

In a similar way, 
\be\label{barcons3a}
\partial_t n_E + \frac{1}{r^2}\partial_r\left[r^2 
n_E \left( N^r + \frac{N}{A}\,^3\,\!U^{(r)}\right)\right]
+  n_E\left[\partial_t \alpha + 2\partial_t \beta 
+ \left( N^r + \frac{N}{A}\,^3\,\!U^{(r)}\right)
 \left(\partial_r \alpha + 2\partial_r \beta\right)\right]
 =0\,\,\,.
\ee

The total contribution of sources [Eqs. (\ref{EPPFT})$-$(\ref{trSPPFT})] 
that appear in the 3+1 Einstein equations 
write in spherical symmetry as follows:
\bea
\label{Etot}
E &=& E_{\rm PF} + E_{\rm R}\,\,\,,\\
J^{(r)} &=& A J^r= (E_{\rm PF} + p )\,U^{(r)} + H_{\rm R}\,\,\,, \\
\label{Srrtot}
S^{(r)}_{\,\,\,\,(r)} &=& (E_{\rm PF} + p )\,(U^{(r)})^2 + p + p_{\rm R}\,\,\,\,,\\
S^{(\theta)}_{\,\,\,\,(\theta)} &=& S^{(\phi)}_{\,\,\,\,(\phi)}= p + p^T_{\rm R} 
 \,\,\,,\\
S &=& S^{(i)}_{\,\,\,\,(i)}= (E_{\rm PF} + p )\,(U^{(r)})^2 + 3p + 
p_{\rm R} + 2 p^T_{\rm R} \,\,\,.
\eea

\section{Gauge and slicing condition}

Two of the most popular gauges and time slicings 
that have been used in spherical symmetry are the {\it radial} and {\it isotropic} 
gauges and the 
{\it maximal} and {\it polar} slicings 
(see Refs. \cite{york79,SmarYork,BarPir} for 
a more general discussion on the slicing choices). Moreover, it is 
in the framework of asymptotically flat spacetimes (condition usually demanded 
in astrophysical applications) that they become specially 
useful; it is in this context that such gauges and time slicings will be 
discussed in the following. 

The isotropic gauge $A=B$ with the maximal 
slicing condition $K=0$, $\partial_t K=0$ (IGMS) has been employed by 
several authors (e.g. see \cite{ST80,STP85,MM,ST85b,ST86,RST}). In the vacuum and static 
case, that choice leads to the well known Schwarzschild solution 
in isotropic coordinates. 
The maximal slicing has the advantage of 
freezing the evolution in regions near the formation of space-like 
singularities while allowing a faster evolution in the outer regions 
(feature usually quoted as ``singularity avoidance'' property). 
The time slicing leads to an elliptic equation for the lapse and therefore, 
for rather general matter conditions (e.g. {\it strong energy condition}), 
one can use the maximum-minimum principle to determine the qualitative 
behavior for the lapse (cf. \cite{york79,SmarYork}). 
The lapse function has a minimum at $r=0$ and a maximum at 
$r\rightarrow \infty$ and during the evolution the minimum tends to zero 
as $t\rightarrow \infty$ 
(the collapse of the lapse) halting the propertime separation between 
neighbouring slices as the singularity forms. However, far from the 
origin $N\rightarrow 1$ which allows to advance the 
evolution in the asymptotic regions. The IGMS coordinates 
have also the advantage that the three-metric remains regular at 
the formation of apparent horizons, which allows to continue the evolution. 
The drawback is that eventually the metric potencial $A$ grows 
exponentially at the origin and then the coordinates are ``sucked down'' 
to the black hole, which avoids a good description of the evolution outside the 
event horizon.

The {\it radial gauge} $B=1$ with the polar slicing condition (RGPS) 
$K= K^{r}_{\,\,\,\,r} $ has been employed sistematically by 
Gourgoulhon \cite{gour1,gour2,gour3}. 
These coordinates are a generalization of the  
Schwarszschild coordinates to the non-static and non-vacuum spacetimes.
The field equations turns to be much more simpler than those of the IGMS 
coordinates since the equations for $A$ and $N$ reduce to 
first order in $r$ while evolving in time  through their sources. Furthermore, the 
shift $N^r$ is zero everywhere on the slices.
 The RGPS coordinates has a central ``singularity avoidance''  
property which is even stronger that the IGMS coordinates. In fact, 
the slowing of the evolution is such that it 
avoids the formation of apparent horizons, the metric 
potential $A$, however, diverges at the star's surface 
as the matter enters 
the Schwarzschild radius \cite{gour1,STP86}, leading, unlike the 
IGMS to a coordinate crash. Thus, these coordinates do not serve to 
describe the black hole interior. 
Nevertheless, the pathological behavior of the RGPS 
coordinates occurs at a large $t$, and 
from the astrophysical point of view (e.g., from the point of 
view of an observer at spatial infinity),
these coordinates 
are good enough to describe the entire evolution of matter outside the black hole, 
the ingoing matter takes by the way, for an observer at infinity, an infinite
time to cross the event horizon (the evolution is thus ``frozen'').

The hybrid choice of isotropic gauge and polar slicing (IGPS) has been less used 
in the past. However, it seems that they overcome the drawbacks of 
the above coordinate choices (cf. \cite{STP86,ST86}; see also 
Ref. \cite{BarPir} for an analysis of these coordinates in the 
context of axisymmetry).

Another popular gauge choice is the 
comovil gauge (Lagrangian gauge). This gauge has been particularly used 
in the study of supernovae collapse with syncronous 
\cite{MW,Lieben2,MS,Baron,Cern} and polar slicings \cite{SBP,Schinder}. The comovil coordinates 
have the advantage that the hydrodynamic equations are simpler since 
there is no advection. The synchronous slicing are orthogonal 
in the sense that there is no shift ($N^r=0$). The disadvantage of the 
latter is that they fail badly when black holes start forming 
(cf. \cite{SBP,Piran}). 
The asynchronous coordinates can remedy this problem. In particular Schinder 
and coauthors \cite{SBP,Schinder}, have used polar slincings 
to avoid the pathologies of the synchronous gauge. Another modification of 
the comovil and syncronous coordinates that allows to handle the 
formation of black holes is the introduction of an outgoing null coordinate 
instead of the usual time coordinate \cite{Baum0,HM}.
\bigskip

Finally, we mention the isotropic gauge and constant-mean-curvature 
slicings ($IGK_t-$slicings) 
$K= K(t)$, employed by Harleston and coauthoros \cite{HV,HH}. This 
time slicing contains the maximal slicig as a particular case. It also posseses 
the feature of strong crushing coordinate avoidance. These coordinates generalize 
(to the non-homogenoues case) 
the comoving coordinates of homogeneous and isotropic spacetimes which are 
relevant in the standard cosmology. Such a choice is thus useful when 
the space-time is required to be asymptotically Friedmann-Robertson-Walker 
\cite{HH,HV}. The main difference between this choice and 
the maximal slicing condition is thus the behavior of the hipersurfaces asymptotically: 
the maximal hypersurfaces reach spatial infinity while the $K_t$-hypersurfaces reach 
future or past null infinity whether $K$ is positive or negative 
\cite{Piran}. 

\subsection{Isotropic gauge and maximal slicing condition (IGMS)}
The isotropic choice $A=B$ ($\alpha=\beta$), 
implies from Eq.(\ref{K^i_je}) that
\be
K^{r}_{\,\,\,\,r}= K^{(r)}_{\,\,\,(r)}= 
-\frac{1}{N}\left(\partial_t \alpha + 
\partial_r N^{r} + N^r \partial_r\alpha \right)\,\,\,.
\ee
Therefore
\be\label{evalfa}
\partial_t \alpha = -N K^{r}_{\,\,\,\,r} - 
\partial_r N^{r} - N^r \partial_r\alpha\,\,\,.
\ee
On the other hand, the maximal slicing condition $K=0$ implies that
\be\label{msc}
K^{r}_{\,\,\,\,r}= -2K^{\theta}_{\,\,\,\,\theta}\,\,\,,
\ee
or equivalently
\be
3\partial_t\alpha + 3N^r\partial_r\alpha +
 \partial_r N^r  + \frac{2N^r}{r} =0\,\,\,.
\ee
Using (\ref{evalfa}) in previous Eq. we obtain
\be\label{pdeNr}
\partial_r N^r - \frac{N^r}{r}= -\frac{3}{2} N K^{r}_{\,\,\,\,r}\,\,\,,
\ee
This can be written as to give the following differential equation for $N^r$
[cf. Eq.(21) of Shapiro \& 
Teukolsky (1980) \cite{ST80}],
\be\label{NrIGMS}
\partial_r \left(\frac{N^r}{r}\right)= -\frac{3}{2r} N K^{r}_{\,\,\,\,r}
\,\,\,\,.
\ee

On the other hand, using Eq. (\ref{pdeNr}) in Eq. (\ref{evalfa}) one 
obtains an evolution equation for $\alpha$:
\be\label{evalfaIGMS}
\partial_t \alpha = -N^r\left(\partial_r\alpha +\frac{1}{r}\right) 
+ \frac{1}{2} N K^{r}_{\,\,\,\,r} \,\,\,.
\ee

With the above choice and with (\ref{msc}), 
the Hamiltonian constraint Eq. (\ref{ham3}) reads,
\be
\label{ham4}
2\,\partial^2_{rr} \alpha + (\partial_r \alpha)^2
  + \frac{4\,\partial_r \alpha}{r}  = 
-8\pi G_0 E A^2 - \frac{3}{4} A^2 (K^{r}_{\,\,\,\,r})^2 \,\,\,\,.
\ee
Adopting the variable
\be
\tilde \alpha= \alpha/2\,\,\,\,,
\ee
we obtain a second order differential equation for $\tilde \alpha$
\be
\label{ham5}
\,\partial^2_{rr} \tilde \alpha 
  + \frac{2\,\partial_r \tilde \alpha}{r}  = -A^2\left[ 
2\pi G_0 E  + \frac{3}{16} (K^{r}_{\,\,\,\,r})^2\right] 
-  (\partial_r \tilde \alpha)^2\,\,\,\,,
\ee
where we recognize in the l.h.s the Laplacian operator of a 
spherically symmetric Euclidean space [cf. 
Eq.(19) of Shapiro \& Teukolsky (1980) \cite{ST80} for a 
source term $E$ including a perfect fluid alone Eq.(\ref{Er})].

The momentum constraint Eq.(\ref{J_r5}), reads
\be\label{J_r6}
3K^{r}_{\,\,\,\,r} 
\left(\frac{1}{r} + \partial_r \alpha \right) + 
 \partial_r K^{r}_{\,\,\,\,r} 
= 8\,\pi G_0 \,J_{r}\,\,\,\,,
\ee 
were we used $J_r= AJ_{(r)}$. This 
can be written as a differental equation for $K^{r}_{\,\,\,\,r}$ as 
[cf. Eq.(20) of Shapiro \& Teukolsky (1980) \cite{ST80}],
\be\label{J_r7}
\partial_r\left(A^3r^3 K^{r}_{\,\,\,\,r}\right)= 8\,\pi G_0 
r^3 A^3 \,J_{r}\,\,\,.
\ee

The Eq.(\ref{dynK4}) provides an elliptic equation for $N$ [cf. 
Eq. (18) of Shapiro \& 
Teukolsky (1980) \cite{ST80} for source terms $E$ and $S$ 
of a perfect fluid Eqs. (\ref{Er}) and (\ref{Srr})],
\be
\label{dynK5}
\partial^2_{rr}\nu + (\partial_r\nu)^2
+(\partial_r \alpha)\,(\partial_r \nu)
+ \frac{2\partial_r \nu}{r} 
=   4\,\pi G_0 \,A^2\,\left(S + E\right) 
+ \frac{3}{2} A^2 (K^{r}_{\,\,\,\,r})^2 \,\,\,\,,
\ee

We have then four differential Eqs. (\ref{NrIGMS}), (\ref{ham5}), 
(\ref{J_r6}) and (\ref{dynK5}) for $N^r$, $A$ and $K^{r}_{\,\,\,\,r}$ 
y $N$ respectively. It is to note in those equations 
that the field variables evolve in time through 
the matter fields. Although the evolution equation for $K^{r}_{\,\,\,\,r}$ 
is redundant, for completness we write it in the IGMS coordinates. 
Equations (\ref{dynk113}), (\ref{ham4}) and (\ref{dynK5}), lead to
\be
 \partial_t K^{(r)}_{\,\,\,\,(r)} + 
{\frac{N^{(r)}\,\partial_r K^{(r)}_{\,\,\,\,(r)}}{A}} +\frac{3}{4}
N \left(K^{(r)}_{\,\,\,\,(r)}\right)^2  - \frac{N}{A^2}\left[ \left(
\partial_r\nu + \partial_r\alpha\right)\left(\frac{2}{r}+\partial_r\alpha\right)
+ (\partial_r\nu)(\partial_r\alpha)\right]
  = - 8\,\pi G_0 \,N\,S^{(r)}_{\,\,\,\,(r)} \,\,\,\,.
\ee

Concerning the energy conservation equation (\ref{ECEfss}), this reads,
\be\label{ECEfigms}
\partial_t E_{\rm PF} +  N^r\partial_r E_{\rm PF}  =  - N\partial_r J^r_{\rm PF} 
- N J^r_{\rm PF} \left(2 \partial_r \nu + \frac{2}{r} + 3\partial_r \alpha\right)  
+ N  K^{(r)}_{\,\,\,\,(r)} \left( \,_{\rm PF}\,\!S^{(r)}_{\,\,\,\,(r)}
- \,_{\rm PF}\,\!S^{(\theta)}_{\,\,\,\,(\theta)}\right)  - N^2 {\cal F}^t
\,\,\,\,.
\ee
Or in terms of $J_r= A^2 J^r$ we obtain
\be\label{ECEfigms2}
\partial_t E_{\rm PF} +  N^r\partial_r E_{\rm PF}  =  - \frac{N}{A^2}\partial_r 
J_r^{\rm PF} - \frac{N}{A^2} J_r^{\rm PF} \left(2 \partial_r \nu + \frac{2}{r} + 
\partial_r \alpha\right)  
+ N  K^{(r)}_{\,\,\,\,(r)} \left(\,_{\rm PF}\,\! S^{(r)}_{\,\,\,\,(r)}
- \,_{\rm PF}\,\!S^{(\theta)}_{\,\,\,\,(\theta)}\right)  - N^2{\cal F}_t
\,\,\,\,.
\ee
The conservative form Eq. (\ref{ECEcons}) writes in this gauge as
\bea
\partial_t E_{\rm PF} + \frac{1}{r^2}\partial_r
\left(r^2 V^r \,E_{\rm PF}\right) &=& 
E_{\rm PF}\left(\frac{2N^r}{r} + \partial_r N^r\right) - 
\frac{1}{r^2}\partial_r
\left( r^2 \frac{N}{A} \,^3\,\!U^{(r)}\,p\right) \nonumber \\
\label{ECEconsIGMS}
&& -\frac{N J^{\rm PF}_{(r)}}{A}\left[ \partial_r\nu + 3\partial_r\alpha 
- \,^3\,\!U^{(r)}\,A K^{(r)}_{\,\,\,\,(r)}  
+ N A {\cal F}^t\right]
\,\,\,\,.
\eea

The momentum conservation equations writes
\bea \label{ECMfigms}
 \partial_t J_{r}^{\rm PF}  +  N^r\partial_r J_r^{\rm PF} = 
- J_r^{\rm PF}\partial_r N^r 
 & & - N\left[ \partial_r \,_{\rm PF}\,\!S^{(r)}_{\,\,\,\,(r)}
 + 2\left( \,_{\rm PF}\,\!S^{(r)}_{\,\,\,\,(r)} - 
\,_{\rm PF}\,\!S^{(\theta)}_{\,\,\,\,(\theta)}\right)\left( \frac{1}{r} + 
\partial_r \alpha  \right)  \right. \nonumber \\
&& \left. +  \left( \,_{\rm PF}\,\!S^{(r)}_{\,\,\,\,(r)} + 
  E_{\rm PF} \right) \partial_r \nu  + \,^3{\cal F}_r\right] \,\,\,.
\eea
In terms of triad components Eq. (\ref{JconsV}) reads,
\bea\label{JconsVIGMS}
 \partial_t \,J^{\rm PF}_{(r)}  + \frac{1}{r^2}\partial_r\left(
r^2 V^r J^{\rm PF}_{(r)}\right) &=& 
J^{\rm PF}_{(r)}\left[ N K^{(r)}_{\,\,\,\,(r)} 
 - 3 V^r \partial_r\alpha 
 + N^r\left(3\partial_r\alpha -\frac{2}{r}\right)
 -  \partial_r N^r\right]
\nonumber \\
&&  - \frac{N}{A}\left[\left(E_{\rm PF}+ p\right)\partial_r\nu 
+\partial_r p + \, ^3{\cal F}_{(r)} A \right]  \,\,\,.
\eea

In this gauge, the Euler equation (\ref{EulerRGr}) reads
\bea \label{EulerIGMS}
& & \partial_{(t)} \,^3\,\!U^{(r)} + \,^3\,\!U^{(r)}\,\,
^3 \partial_{(r)}\,^3\,\!U^{(r)}  = 
- \frac{1}{E_{\rm PF}+p} \left[ \,^3\partial^{(r)}p + \,^3\,\!U^{(r)} \partial_{(t)}p 
\right]  \nonumber \\
\!\!\!\!\!\!\!\!\!\!\!\!
&+& \frac{1}{\Gamma^2}\left(\,\,^3\,\!U^{(r)} K^{(r)}_{\,\,\,(r)}- 
\,^3\partial_{(r)}\nu\,\right) 
+ \frac{1}{E_{\rm PF}+p}\left(\,^3\,\!U^{(r)}{\cal F}^{(t)} - {\cal F}^{(r)}\right)
\,\,\,\,.
\eea

Under the IGMS coordinates the evolution equations for the 
entropy per baryon and the particle number per baryon keep the 
same form as Eqs. (\ref{entconscoorr}) and (\ref{partconscoorr})
or the arternative form given by Eqs. 
(\ref{entcons4}) and (\ref{partcons4}), where as 
Eqs. (\ref{barcons1r}) and (\ref{barcons3}) read

\be\label{barcons1IGMS}
\partial_t \left(A^3 n_E\right) + \frac{1}{r^2}\partial_r\left[r^2 
 V^r n_E A^3\right] =0\,\,\,,
\ee

\be\label{barconsIGMS}
\partial_t \left(A^3 n_E\right) + \frac{1}{r^2}\partial_r\left[A^3 r^2 
n_E \left( N^r + \frac{N}{A}\,^3\,\!U^{(r)}\right)\right] =0\,\,\,.
\ee

The equation (\ref{barcons1IGMS}) has a conservative form for the 
quantity $n_{\rm E} A^3$.

When using the evolution equation (\ref{evalfaIGMS}), the Eqs. 
(\ref{barcons1IGMS}) and (\ref{barconsIGMS}) become respectively 

\be\label{barcons1IGMS2}
\partial_t n_{\rm E} + \frac{1}{r^2}\partial_r\left[r^2 V^r  n_{\rm E}\right]
+ 3n_{\rm E}\left[ \left(V^r - N^r\right)\partial_r \alpha - 
\frac{N^r}{r} + \frac{1}{2} N K^{r}_{\,\,\,\,r}\right] =0\,\,\,,
\ee

\be\label{barconsIGMS2}
\partial_t n_{\rm E} + \frac{1}{r^2}\partial_r\left[r^2 n_{\rm E} \left( N^r +
\frac{N}{A}\,^3\,\!U^{(r)}\right) \right]
+ 3n_{\rm E}\left[\frac{N}{A}\,^3\,\!U^{(r)} \partial_r \alpha - 
\frac{N^r}{r} + \frac{1}{2} N K^{r}_{\,\,\,\,r}\right] =0\,\,\,.
\ee

The equation (\ref{barcons1IGMS2}) has a conservative 
form for the quantity $n_E$.

\subsection{Radial gauge and polar slicing condition (RGPS)}
In this gauge, $B=1$ ($\beta=0$) and the polar slicing condition 
$K= K^{r}_{\,\,\,\,r}$ is equivalent to  $K^{\theta}_{\,\,\,\,\theta}
+ K^{\phi}_{\,\,\,\,\phi}=0$. Since $K^{\theta}_{\,\,\,\,\theta}= 
K^{\phi}_{\,\,\,\,\phi}$, this slicing condition and 
the gauge choice lead to $N^r=0$. 

The Hamiltonian constraint Eq.(\ref{ham3}) reads
\be\label{hamrgps}
\frac{1}{r^2}\left(A^2 -1\right)   
  + \frac{2}{r}\partial_r \alpha = 8\pi G_0 E A^2 \,\,\,\,.
\ee
Moreover, by defining
\be\label{adef}
A(r,t):= \left( 1- \frac{2G_0 m(r,t)}{r}\right)^{-1/2}\,\,\,\,,
\ee
the Eq.(\ref{hamrgps}) reads,
\be\label{rgpsA}
\partial_r \alpha = A^2\,G_0\left( 4\pi r E -\frac{m}{r^2}\right)
\ee
or even [cf. Eq.(18) of Ref. \cite{gour1} or Eq.(3.29) of Ref. \cite{gour3} 
for a perfect fluid alone or for a perfect fluid accompanied by a neutrino flow, 
respectively],
\be\label{hamrgps2}
\partial_r m = 4\pi r^2 E\,\,\,\,.
\ee

The momentum constraint (\ref{J_r4}), for the present gauge 
choice reads
\be\label{J_rrgps}
K^{(r)}_{\,\,\,\,(r)}= -\frac{1}{AN}\partial_t A
= 4\,\pi r G_0 \,J_{r}=  4\,\pi r G_0\, A^2 \,J^{r}= 
4\,\pi r G_0\, A \,J^{(r)}   \,\,\,\,,
\ee 
This with Eq. (\ref{adef}) results in an evolution equation for $m(r,t)$ 
[cf. Eq.(20) of Ref. \cite{gour1} or Eq.(3.31) of Ref. \cite{gour3} 
for a perfect fluid alone or for a perfect fluid accompanied by a neutrino flow, 
respectively]:
\be\label{evm}
\partial_t m= -4\pi r^2 N J^r\,\,\,\,.
\ee

The evolution equation (\ref{dynk223}) gives,
\be\label{rgpsnu}
\frac{\partial_r \nu}{r} - \left[ \frac{1}{r^2}\left(A^2-1\right) 
+ \frac{\partial_r A}{rA} \right]  
=   4\,\pi G_0\,A^2 \,\left(S^{(r)}_{\,\,\,\,(r)}  - E\right) \,\,\,\,.
\ee
With (\ref{adef}) and (\ref{hamrgps}) this writes [cf. 
Eq.(22) of Ref. \cite{gour1} and Eq.(3.32) of Ref. \cite{gour3} 
for a source term $S^{(r)}_{\,\,\,\,(r)}$ of a perfect fluid alone 
Eq. (\ref{Srr}) or that of a perfect fluid accompanied by a neutrino 
flow (\ref{Srrtot}), respectively], 
\be\label{rgpsnu2}
\partial_r \nu 
=   G_0 \,A^2 \,\left(\frac{m}{r^2}+ 4\pi\,r\,
S^{(r)}_{\,\,\,\,(r)} \right) \,\,\,\,.
\ee

Therefore Eqs. (\ref{hamrgps2}) and (\ref{rgpsnu2}) are the field equations 
for the two variables $A$ and $N$ respectively. These quantities evolve in 
time through the matter variables. Therefore, as in the IGMS 
coordinate choice, the evolution equation for $K^{(r)}_{\,\,\,\,(r)}$ 
is also redundant in the RGPS coordinates. For completeness we write it 
using Eq. (\ref{dynk113}) [cf. Eq.(21) 
of Ref. \cite{gour1} for a perfect fluid alone],  
\be
 \partial_t K^{(r)}_{\,\,\,\,(r)}  -
N \left(K^{(r)}_{\,\,\,\,(r)}\right)^2  - \frac{N}{A^2}\left[
 \frac{2\partial_r \alpha}{r}  
 - \partial^2_{rr}\nu + (\partial_r \nu) \left(\partial_r\alpha -
\partial_r \nu\right)  \right]    
 =  4\,\pi G_0 \,N\,\left( -S^{(r)}_{\,\,\,\,(r)} + 
  2 S^{(\theta)}_{\,\,\,\,(\theta)} - E\right) \,\,\,\,.
\ee
Concerning the evolution equations for the matter, Eq.(\ref{ECEfss}) reads,
\be\label{ECEfrgps}
\partial_t E_{\rm PF} + \frac{N}{Ar^2}\partial_r
\left(Ar^2 J^r_{\rm PF} \right) = 
N  K^{(r)}_{\,\,\,\,(r)} \left(  \,_{\rm PF}\,\!S^{(r)}_{\,\,\,\,(r)}  
+  E_{\rm PF} \right) - 2 J^r_{\rm PF}  \partial_r N  - N^2 {\cal F}^t
\,\,\,\,,
\ee
which can be written as
\be\label{ECEfrgps2}
\partial_t E_{\rm PF} + \frac{1}{r^2}\partial_r
\left(Nr^2 J^r_{\rm PF} \right) = 
N  K^{(r)}_{\,\,\,\,(r)} \left(  \,_{\rm PF}\,\!S^{(r)}_{\,\,\,\,(r)}  
+  E_{\rm PF} \right) - N J^r_{\rm PF} \left( \partial_r \nu + \partial_r\alpha\right)
- N^2 {\cal F}^t
\,\,\,\,.
\ee
We can replace the gradients of the metric potentials by using
the Eqs. (\ref{hamrgps}) and (\ref{rgpsnu}) which imply,
\be\label{dnudalfa}
\partial_r \nu + \partial_r\alpha= 
4\,\pi r G_0\, A^2 \left( S^{(r)}_{\,\,\,\,(r)}  
+  E \right)\,\,\,.
\ee
Then
\be
\partial_t E_{\rm PF} + \frac{1}{r^2}\partial_r
\left(Nr^2 J^r_{\rm PF} \right) = 
N  K^{(r)}_{\,\,\,\,(r)} \left(  \,_{\rm PF}\,\!S^{(r)}_{\,\,\,\,(r)}  
+  E_{\rm PF} \right) -   
4\,\pi r G_0\, N A^2 J^r_{\rm PF} \left( S^{(r)}_{\,\,\,\,(r)}  
+  E \right) - N^2 {\cal F}^t
\,\,\,\,.
\ee

Using the momentum constraint (\ref{J_rrgps}), we obtain
\be\label{ECEfrgps3}
\partial_t E_{\rm PF} + \frac{1}{r^2}\partial_r
\left(Nr^2 J^r_{\rm PF} \right) = 4\,\pi r G_0\, N A^2\left[
J^{r} \left(  \,_{\rm PF}\,\!S^{(r)}_{\,\,\,\,(r)}  
+  E_{\rm PF} \right) -  J^r_{\rm PF} \left( S^{(r)}_{\,\,\,\,(r)}  
+  E \right)\right] - N^2 {\cal F}^t
\,\,\,\,.
\ee
Now, since
\bea
\label{Jrtot}
J^r &=&  J^r_{\rm PF} +  J^r_{\rm R} \,\,\,\,\,,\\
E &=& E_{\rm PF} + E_{\rm R}\,\,\,,\\
\label{Srrtot2}
S^{(r)}_{\,\,\,\,(r)} &=& \,_{\rm PF}\,\!S^{(r)}_{\,\,\,\,(r)} 
+ \,_{\rm R}\,\!S^{(r)}_{\,\,\,\,(r)} \,\,\,\,\,\,,
\eea
We find
\be\label{ECEfrgps4}
\partial_t E_{\rm PF} + \frac{1}{r^2}\partial_r
\left(Nr^2 J^r_{\rm PF} \right) = 4\,\pi r G_0\, N A^2\left[
J^{r}_{\rm R} \left(  \,_{\rm PF}\,\!S^{(r)}_{\,\,\,\,(r)}  
+  E_{\rm PF} \right) -  J^r_{\rm PF} 
\left(  \,_{\rm R}\,\!S^{(r)}_{\,\,\,\,(r)}  
+  E_{\rm R} \right) \right]
- N^2 {\cal F}^t \,\,\,\,.
\ee
Using that $J^{(r)}_{\rm PF}= A J^r_{\rm PF}$,  $J^{(r)}_{\rm R}= A J^r_{\rm R}$ 
and Eqs. (\ref{Srr}) and (\ref{JPPFr}), 
the energy conservation Eq.(\ref{ECEfrgps4}) finally reads 
[cf. Eq.(25) of Ref. \cite{gour1} or Eq.(3.55) of Ref. \cite{gour3} 
for a perfect fluid alone or for a perfect fluid accompanied by a neutrino flow, 
respectively],
\be\label{ECEfrgps5}
\partial_t E_{\rm PF} + \frac{1}{r^2}\partial_r\left[r^2(E_{\rm PF}+ p)V^r\right]= 
4\,\pi r G_0\,N A\,\left(E_{\rm PF} + p\right)\left[ 
J^{(r)}_{\rm R}\left( (U^{(r)})^2 + 1\right) - U^{(r)} 
\left(  \,_{\rm R}\,\!S^{(r)}_{\,\,\,\,(r)}  
+  E_{\rm R} \right) \right] - N^2{\cal F}^t
\,\,\,\,.
\ee 

The alternative expression of Eq. (\ref{ECEfrgps5}) in conservative form reads
\bea\label{ECEconsRGPS}
\partial_t E_{\rm PF} + \frac{1}{r^2}\partial_r\left(r^2 V^r E_{\rm PF} \right) &=& 
- \frac{1}{r^2}\partial_r\left(r^2 V^r\,p\right)
 + 4\,\pi r G_0\,N A\,\left(E_{\rm PF} + p\right)\left[ 
J^{(r)}_{\rm R}\left( (U^{(r)})^2 + 1\right) - U^{(r)} 
\left(  \,_{\rm R}\,\!S^{(r)}_{\,\,\,\,(r)}  
+  E_{\rm R} \right) \right] \nonumber \\
&& - N^2{\cal F}^t
\,\,\,\,.
\eea 

The momentum conservation Eq.(\ref{ECMfss2}) reads
\be \label{ECMfrgps}
 \partial_t J_{r}^{\rm PF}  + N\,\partial_r \,_{\rm PF}\,\! S^{(r)}_{\,\,\,\,(r)}
 + \frac{2N}{r} \,_{\rm PF}\,\!\left(S^{(r)}_{\,\,\,\,(r)} - 
\,_{\rm PF}\,\!S^{(\theta)}_{\,\,\,\,(\theta)}\right) 
  = -\left( \,_{\rm PF}\,\!S^{(r)}_{\,\,\,\,(r)} + 
  E_{\rm PF}\right) \partial_r N  + N\,J_{r}^{\rm PF}  K^{(r)}_{\,\,\,\,(r)} 
- \, ^3{\cal F}_r N \,\,\,.
\ee
This can be written as
\be \label{ECMfrgps2}
 \partial_t J_{r}^{\rm PF} =  N\left[ -\frac{1}{r^2N}\partial_r
\left(r^2 N  \,_{\rm PF}\,\!S^{(r)}_{\,\,\,\,(r)}\right)
 + \frac{2}{r} \,_{\rm PF}\,\!S^{(\theta)}_{\,\,\,\,(\theta)}  
  -E_{\rm PF} \partial_r \nu  + J_{r}^{\rm PF} K^{(r)}_{\,\,\,\,(r)}
- \, ^3{\cal F}_r \right] \,\,\,.
\ee

Using (\ref{JconsV}) the conservative form of this latter reads 
\be\label{JconsVRGPS}
 \partial_t \,J^{\rm PF}_{(r)}  + \frac{1}{r^2}\partial_r\left(
r^2 V^r J^{\rm PF}_{(r)}\right) = 
J^{\rm PF}_{(r)}\left[ 2N K^{(r)}_{\,\,\,\,(r)} 
   - V^r \partial_r\alpha \right]  
- \frac{N}{A}\left[\left(E_{\rm PF}+ p\right)\partial_r\nu 
+\partial_r p + \, ^3{\cal F}_{(r)} A \right]  \,\,\,.
\ee
Furthermore, the use of Eqs. (\ref{adef}), (\ref{rgpsA}), 
(\ref{J_rrgps}) and (\ref{rgpsnu2}) lead to 
\bea\label{JconsVRGPS2}
 \partial_t \,J^{\rm PF}_{(r)}  + \frac{1}{r^2}\partial_r\left(
r^2 V^r J^{\rm PF}_{(r)}\right) &=& G_0\,
J^{\rm PF}_{(r)}\left[ 8\pi r A N J^{(r)} 
-  A^2 V^r \left(4\pi r E - \frac{m}{r^2}\right) \right]  \nonumber \\
&&
- G_0 N A \left(E_{\rm PF}+ p\right)\left(\frac{m}{r^2} + 4\pi r 
S^{(r)}_{\,\,\,\,(r)}\right)
-\frac{N}{A}\left(\partial_r p + \, ^3{\cal F}_{(r)} A \right)  \,\,\,.
\eea
Finally, using Eqs. (\ref{Jrtot}) and (\ref{Srrtot2}) and the 
expressions (\ref{Srr}) and (\ref{JPPFr}), we obtain
\bea\label{JconsVRGPS3}
 \partial_t \,J^{\rm PF}_{(r)}  + \frac{1}{r^2}\partial_r\left(
r^2 V^r J^{\rm PF}_{(r)}\right) &=& G_0\,
J^{\rm PF}_{(r)}\left\{ 4\pi r A N \left(J^{\rm PF}_{(r)} + 2J^{(r)}_{\rm R}\right) 
-  A^2 V^r \left[4\pi r \left(E_{\rm PF} + E_{\rm R}\right) 
- \frac{m}{r^2}\right] \right\}  \nonumber \\
&&
- G_0 N A \left(E_{\rm PF}+ p\right)\left[\frac{m}{r^2} + 4\pi r\left(p 
+ \,_{\rm R}\,\!S^{(r)}_{\,\,\,\,(r)} \right)\right]
-\frac{N}{A}\left(\partial_r p + \, ^3{\cal F}_{(r)} A \right)  \,\,\,.
\eea

The Euler equation (\ref{EulerRGr}) and Eqs. (\ref{J_rrgps}) and 
(\ref{rgpsnu2}) lead to
\bea \label{Eulerfrgps4}
& & \partial_{(t)} \,^3\,\!U^{(r)} + \,^3\,\!U^{(r)}\,\,
^3 \partial_{(r)}\,^3\,\!U^{(r)}  = 
- \frac{1}{E_{\rm PF}+p} \left[ \,^3\partial^{(r)}p + \,^3\,\!U^{(r)} \partial_{(t)}p 
\right]  \nonumber \\
\!\!\!\!\!\!\!\!\!\!\!\!
&-& \frac{G_0 A}{\Gamma^2}\left[\frac{m}{r^2} + 4\pi\,r\left(S^{(r)}_{\,\,\,\,(r)}
-\,\,^3\,\!U^{(r)} J_{(r)}\right)\right] 
+ \frac{1}{E_{\rm PF}+p}\left(\,^3\,\!U^{(r)}{\cal F}^{(t)} - \,^3\,\!{\cal F}^{(r)}\right)
\,\,\,\,.
\eea

Again, using Eqs. (\ref{Jrtot}) and (\ref{Srrtot2}) and the 
expressions (\ref{Srr}) and (\ref{JPPFr}), we obtain
\bea\label{ECMfrgps5}
\partial_t\,^3\,\! U^{(r)} + V^r
\partial_r\,^3\,\! U^{(r)} &=&  -\frac{1}{(E_{\rm PF}+ p)}\left(
\frac{N}{A}\partial_r p
+ \,^3\,\!U^{(r)}\partial_t p \right) -\frac{ANG_0}{\Gamma ^2}\left[ \frac{m}{r^2}
+ 4\pi r \left( p + \,_{\rm R}\,\!S^{(r)}_{\,\,\,\,(r)}- \,^3\,\!U^{(r)}
J^{(r)}_{\rm R} \right) \right] \nonumber \\
&& +\frac{1}{(E_{\rm PF}+ p)}\left(\,^3\,\!U^{(r)}{\cal F}^{(t)} -\, ^3{\cal F}^{(r)} N\right)
\eea
where we also used that in this gauge $\partial_{(t)}= 1/N \partial_t$, 
$\partial_{(r)}= 1/A\partial_r$ and  $U^{(r)}= \frac{A}{N} V^r$.

This is the Euler equation of the fluid in spherical symmetry 
which includes the forces of the radiation fields acting on the fluid  
[cf. Eq.(34) of Ref. \cite{gour1} or Eq.(3.56) of Ref. \cite{gour3} 
for a perfect fluid alone or for a perfect fluid accompanied by a neutrino flow, 
respectively]. 
 
A relation that turns to be useful in this gauge is obtained 
by combining Eqs. (\ref{J_rrgps}) and (\ref{dnudalfa}) 
\be\label{AN}
\frac{1}{AN}\partial_t (A^2) + \partial_r \nu + \partial_r \alpha
= 4\pi r G_0 A^2 \left[ S^{(r)}_{\,\,\,\,(r)} + E - 2 J^{(r)}\right]\,\,\,.
\ee
For example, in the static case and for perfect fluids, this provides 
a simple relation between the gradients of the metric potentials and the 
pressure and energy-density of matter. Moreover, outside the star surface the 
only contributions to the total matter variabels are those of radiated 
particles. Under the free streaming 
approximation [cf. Eq.(\ref{free})], this implies that 
the r.h.s of (\ref{AN}) vanishes. This situation was investigated 
analitically in the past using a different gauge \cite{Vaidya,Lind65} 
and corresponds to the external solution. 

Finally, the integrability condition
$\partial^2_{rt} m= \partial^2_{tr} m$ imposed in Eqs. 
(\ref{hamrgps2}) and (\ref{evm}) result in the relationship 
\be\label{evE}
\partial_t E + \frac{1}{r^2}\partial_r\left(r^2 N J^r\right)=0\,\,\,.
\ee
This equation is in fact compatible with the evolution equation for 
the total energy density of matter.

Indeed, subtracting Eq. (\ref{ECEfrgps3}) or more specifically Eq. (\ref{ECEfrgps5}) 
from (\ref{evE}) one obtains an evolution equation for the energy-density of 
radiation $E_{\rm R}$ [cf. Eq. (\ref{ECEftetrad}) ].

The evolution equation (\ref{barcons1}) and the 
alternative form (\ref{barcons3}) in RGPS coordinates read respectively

\be\label{barcons1RGPS}
\partial_t \left(A n_{\rm E}\right) + \frac{1}{r^2}\partial_r\left[r^2 V^r 
A n_{\rm E} \right] =0\,\,\,,
\ee
 
\be\label{barconsRGPS}
\partial_t \left( A n_{\rm E}\right) + 
\frac{1}{r^2}\partial_r\left(r^2 n_{\rm E} N \,^3\,\!U^{(r)}\right) =0\,\,\,.
\ee

Note that Eq. (\ref{barcons1RGPS}) has a conservative form for the 
quantity $A n_{\rm E}$.

Using Eqs. (\ref{J_rrgps}), (\ref{rgpsA}), (\ref{Jrtot}), 
(\ref{JPPFr}) and (\ref{P3Ur}) in Eq. (\ref{barcons1RGPS}) it turns
 
\be\label{barcons1RGPS2}
\partial_t n_{\rm E} + \frac{1}{r^2}\partial_r\left[ r^2 
n_{\rm E} V^r\right] - G_0 A^2 n_{\rm E}
\left[ V^r\left( \frac{m}{r^2} + 4\pi r p\right) + 4\pi r N J^r_{\rm R}
\right]
 =0\,\,\,,
\ee
which provides an equation in conservative form for $n_{\rm E}$.

In the same way, using Eqs. (\ref{J_rrgps}) and (\ref{rgpsnu2}) in 
the alternative Eq. (\ref{barconsRGPS}) we obtain

\be\label{barconsRGPS2}
\partial_t n_{\rm E} + 
\frac{N}{Ar^2}\partial_r\left(r^2 \,^3\,\!U^{(r)} n_{\rm E}\right) 
+ n_{\rm E} G_0 A N \left[ -4\pi r J^{(r)} + \,^3\,\!U^{(r)}\left(
\frac{m}{r^2} + 4\pi r S^{(r)}_{\,\,\,\,(r)}\right)\right]  =0
\ee

The evolution Eqs. (\ref{entconscoorr}) and (\ref{partconscoorr}) 
do not change in form under the RGPS coordinates. However, the 
alternative form given by Eqs. (\ref{entcons4}) and (\ref{partcons4}) in 
RGPS coordinates write respectively
\be\label{entconsRGPS}
\partial_t \sigma + \frac{N \,^3\,\!U^{(r)}}{A} 
\partial_r \sigma= -\frac{N}{n\Gamma \Theta}\left( \mu^{\rm R} {\cal R}_{\rm R}
+ {\cal D}_p\right)\,\,\,.
\ee

\be\label{partconsRGPS}
\partial_t x_{\rm R}  + \frac{N \,^3\,\!U^{(r)}}{A} 
\partial_r x_{\rm R}
= \frac{N}{n\Gamma}{\cal R}_{\rm R}\,\,\,.
\ee

\subsection{Isotropic gauge and polar slicing condition (IGPS)}
The isotropic choice $A=B$ (i.e., $\alpha=\beta$), and the 
polar slicing condition $K^{\theta}_{\,\,\,\,\theta}
+ K^{\phi}_{\,\,\,\,\phi}=0$, implies due to 
the spherical symmetry that $K^{\theta}_{\,\,\,\,\theta}=0$. This latter 
leads to an evolution equation for $\alpha$:
\be\label{alfaIGPS}
\partial_t \alpha= - N^r \partial_r\alpha -\frac{N^r}{r}\,\,\,.
\ee
This and the expression for $K^{r}_{\,\,\,\,r}$ [cf. Eq.(\ref{K^i_je})] 
provide an equation for the 
shift:
\be\label{NrIGPS}
\partial_r\left(\frac{N^r}{r}\right)= -\frac{N}{r} K^{(r)}_{\,\,\,\,(r)}\,\,\,.
\ee
The Hamiltonian constraint (\ref{ham3}) leads to
\be
\label{hamIGPS}
\,\partial^2_{rr} \tilde \alpha 
  + \frac{2\,\partial_r \tilde \alpha}{r}  = 
-2\pi G_0 A^2 E -  (\partial_r \tilde \alpha)^2\,\,\,\,,
\ee
where,
\be
\tilde \alpha= \alpha/2\,\,\,\,.
\ee

The momentum constraint (\ref{J_r4}) writes,
\be
\label{J_rIGPS}
K^{(r)}_{\,\,\,\,(r)} 
\left(\frac{1}{r} + \partial_r \alpha \right) 
= 4\,\pi G_0 \,A\,J_{(r)}= 4\,\pi G_0 J_{r} \,\,\,\,.
\ee 

The equation (\ref{dynk223}) together with (\ref{hamIGPS}) provide 
an equation for the lapse,
\be\label{nuigps}
\partial_r\nu\left(\frac{1}{r} + 2\partial_r\tilde\alpha\right)
= 4\,\pi G_0 A^2 S^{(r)}_{\,\,\,\,(r)} - 
2\partial_r\tilde\alpha \left(\frac{1}{r} + \partial_r\tilde\alpha\right)
\,\,\,\,.
\ee
In the case of a perfect fluid alone, this corresponds to 
Eq. (9) of Shapiro \& Teukolsky \cite{ST86}, and to 
Eq.(6) of Schinder et al.\cite{SBP} where the authors 
use a ``Lagrangian'' gauge which can be easily transformed to 
the IGPS coordinates.

Finally, the evolution equation for $K^{(r)}_{\,\,\,\,(r)}$ given by Eq. 
(\ref{dynk113}) together with Eq. (\ref{hamIGPS}) yield
\bea
 \partial_t K^{(r)}_{\,\,\,\,(r)} + 
{\frac{N^{(r)}\,\partial_r K^{(r)}_{\,\,\,\,(r)}}{A}} -
N \left(K^{(r)}_{\,\,\,\,(r)}\right)^2 &-& \frac{N}{A^2}\left[
- \partial^2_{rr}\nu + (\partial_r \nu) \left(\partial_r\alpha -
\partial_r \nu\right)
+ (\partial_r \alpha) \left(\frac{2}{r} + 2\partial_r\alpha\right) 
\right]\nonumber \\
&= & 4\,\pi G_0 \,N\,\left( -S^{(r)}_{\,\,\,\,(r)} + 
  2 S^{(\theta)}_{\,\,\,\,(\theta)} + E\right) \,\,\,\,.
\eea

The equation of conservation of energy (\ref{ECEcons}) reads in this 
gauge as follows,
\bea
\partial_t E_{\rm PF} + \frac{1}{r^2}\partial_r
\left(r^2 V^r \,E_{\rm PF}\right) &=& E_{\rm PF}\left( N K^{(r)}_{\,\,\,\,(r)} + 
\frac{2N^r}{r} + \partial_r N^r\right) - 
\frac{1}{r^2}\partial_r
\left( r^2 \frac{N}{A} \,^3\,\!U^{(r)}\,p\right) \nonumber \\
\label{ECEconsIGPS}
&& -\frac{N J^{\rm PF}_{(r)}}{A}\left[ \partial_r\nu + 3\partial_r\alpha  
 - \,^3\,\!U^{(r)}\,A K^{(r)}_{\,\,\,\,(r)} \right] 
+ N K^{(r)}_{\,\,\,\,(r)} p  - N^2 {\cal F}^t
\,\,\,\,.
\eea

The equation of conservation of momentum Eq. (\ref{JconsV}) reads
\bea\label{JconsVIGPS}
 \partial_t \,J^{\rm PF}_{(r)}  + \frac{1}{r^2}\partial_r\left(
r^2 V^r J^{\rm PF}_{(r)}\right) &=& 
J^{\rm PF}_{(r)}\left[ 2N K^{(r)}_{\,\,\,\,(r)} 
 - 3V^r \partial_r\alpha +
 + N^r\left(3\partial_r\alpha  -\frac{2}{r}\right)
 -  \partial_r N^r\right]
\nonumber \\
&&  - \frac{N}{A}\left[\left(E_{\rm PF}+ p\right)\partial_r\nu 
+\partial_r p + \, ^3{\cal F}_{(r)} A \right]  \,\,\,.
\eea

The Euler equation (\ref{EulerRGr}) in this gauge take the same form of 
equation (\ref{EulerIGMS}), and the Eqs. (\ref{entcons4}), 
(\ref{partcons4}) and (\ref{barcons3}) [see Eqs. 
(\ref{barcons1IGMS}) and (\ref{barconsIGMS})] 
keep also the same form as in the IGMS coordinates. Altenatively, one 
can also use the simpler form of Eqs. (\ref{entconscoorr}) and 
(\ref{partconscoorr}). Furthermore, when using the evolution 
equation (\ref{alfaIGPS}) in Eqs. (\ref{barcons1r2}) and (\ref{barcons1r2}) 
in IGPS coordinates we obtain respectively

\be\label{barconsIGPS}
\partial_t n_E + \frac{1}{r^2}\partial_r\left[r^2 V^r  n_E\right]
+ 3n_E\left[ \left(V^r - N^r\right)\partial_r \alpha - 
\frac{N^r}{r} \right] =0\,\,\,,
\ee

\be\label{barconsIGPS2}
\partial_t n_E + \frac{1}{r^2}\partial_r\left[r^2 n_E \left( N^r +
\frac{N}{A}\,^3\,\!U^{(r)}\right) \right]
+ 3n_E\left[\frac{N}{A}\,^3\,\!U^{(r)} \partial_r \alpha - 
\frac{N^r}{r}\right] =0\,\,\,.
\ee

\subsection{Boundary conditions and initial data}
A typical feature of spherically symmetric spacetimes is that 
the gravitational field variables can  
evolve in time through the matter fields. 
So the initial conditions for the matter 
variables and the boundary conditions fix automatically the initial 
values for the gravitational field by solving the constraint equations 
and the slicing condition equation. For spacetimes with less symmetries one 
is always forced to solve the dynamic Einstein equations to evolve the 
gravitational field. Let us thus discuss first the 
boundary conditions.

We call {\it exterior solution} the solution of field equations outside the 
perfect-fluid domain (usually a compact support). In the present 
case, it does not correspond to the Schwarzschild vacuum solution since 
in general, the radiated matter will extend to spatial infinity. 
Thus, the exterior solution has to be found also numerically.
The exterior sources of the field equations will be provided by the energy-momentum 
tensor of particles [cf. Eq. (\ref{EMNtensor})]. 
The matter variables of particles will evolve in time 
through the distribution function. Moreover, outside the star, the radiated 
particles can interact only with themselves, however, this interaction is 
rather week in comparison with the interaction inside the star. In 
a first approximation one can thus neglect such interactions and 
consider that the particles will 
follow geodesics; the distribution function will thus remain constant 
along them.

Regarding the boundary conditions, these are rather regularity and 
asymptotic conditions. For instance, the regularity and 
the asymptotic flatness condition for the shift are respectively 
(see Ref. \cite{BarPir} for a more detailed analysis about regularity 
and boundary conditions)
\bea
N^r(t,0) &= &0 \,\,\,\,,\\
N^r(t, r)_{r\rightarrow \infty} &\rightarrow& 0 \,\,\,\,.
\eea
Similar conditions apply for $K^r_{\,\,\,\,r}$. These boundary conditions 
are enforced from Eqs. (\ref{NrIGMS}) and  (\ref{NrIGPS}):
\be
N^r(t,r)= \lambda_{\rm TS}\, r \int_r^\infty \frac{N(t,r')}{r'} K^{(r)}_{\,\,\,\,(r)} (t,r') dr'
\ee
where $\lambda_{\rm TS}= 1,3/2$ for the IGPS and IGMS coordinates respectively.

The condition for the lapse at the star's center is such that 
the asymptotic flatness condition $N\rightarrow 1$ is verified. Therefore 
\be
N(t,0)= N_c(t)\,\,\,,
\ee
with $N_c(t)$ such that 
\be\label{Nasym}
N(t,r)_{r\rightarrow \infty} \rightarrow 1\,\,\,\,.
\ee
Since apriori this is difficult to enforce, a better strategy consists in 
rescaling the 
lapse as $\tilde N= N/N_c$ so that 
$\tilde N_c(t)\equiv 1$, then the values of the true lapse 
can be recovered by using $N_c(t)= 1/\tilde N(t,r)_{r\rightarrow \infty}$ 
where the asymptotic value $\tilde N_\infty(t):= \tilde N(t,r)_{r\rightarrow \infty}$ 
is found numerically at every time 
step. This rescaling allows to integrate the equations spatially in only 
one cycle. The rescaling will not affect the relevant equations of 
motion for $N$ or $A$ since only the derivatives of 
$\nu$ appear there [i.e., the Eqs. (\ref{ham5}), (\ref{dynK5}), 
(\ref{hamrgps}), (\ref{rgpsnu}), (\ref{hamIGPS}), (\ref{nuigps})  
for $N$ and $A$ in the different gauges are invariant to such a rescaling]. 
However, this is not true for the shift equation 
and for the Boltzmann equation where $N$ appears explicitly. 
However, this does not pose  
a problem since a simultaneous rescaling $\tilde N^r= N^r/ N_c$ 
leaves all equations invariant as well as 
the boundary conditions for $\tilde N^r$.

In the case of the RGPS and IGPS coordinates one can find 
an integral expression for 
the lapse satisfying the boundary conditions. For instance, from Eq. (\ref{rgpsnu2}) 
\be\label{rgpsnuint}
\nu(t,r) = G_0 \int_{0}^r \,A^2 \,\left(\frac{m}{r'^2}+ 4\pi\,r'\,
S^{(r)}_{\,\,\,\,(r)} \right) dr' + \nu(t,0) \,\,\,.
\ee
The asymptotic flatness condition Eq.(\ref{Nasym}) leads to
\be
\nu(t,r)_{r\rightarrow \infty} \rightarrow 0 \,\,\,.
\ee
Therefore from Eq. (\ref{rgpsnuint}) and the asymptotic condition one obtains
\be\label{nu0}
\nu(t,0)=  -G_0 \int_{0}^\infty \,A^2 \,\left(\frac{m}{r'^2}+ 4\pi\,r'\,
S^{(r)}_{\,\,\,\,(r)} \right) dr' \,\,\,.
\ee
This corresponds precisely to the renormalized value $-\tilde \nu_\infty$. 
So finally,
\be\label{rgpsnuint2}
 \nu(t,r) =  - G_0 \int_{r}^\infty \,A^2 \,\left(\frac{m}{r'^2}+ 4\pi\,r'\,\,
S^{(r)}_{\,\,\,\,(r)} \right) dr'  \,\,\,\,.
\ee
The value for the lapse at the star surface will be provided by
\be
 \nu(t,R(t)) =  - G_0 \int_{R(t)}^\infty \,A^2 \,\left(\frac{m}{r'^2}+ 4\pi\,r'\,
S^{(r)}_{\,\,\,\,(r)} \right) dr'  \,\,\,\,,
\ee
where $R(t)$ corresponds to the RGPS-r coordinate at the star surface at time 
$t$.  We emphasize that, outside the star 
$S^{(r)}_{\,\,\,\,(r)}= \,_{\rm R}\,\!S^{(r)}_{\,\,\,\,(r)}$, that is, 
the only contribution to $S^{(r)}_{\,\,\,\,(r)}$ is from 
the radiated particles. In fact, outside the star we can write,

\bea
\nu(t,r)_{\rm out} &=& -G_0 \int_{r\geq R(t)}^\infty \,A^2 \,\left(\frac{m}{r'^2}+ 4\pi\,r'\,
_{\rm R}\,\!S^{(r)}_{\,\,\,\,(r)} \right) dr' \nonumber \\
 &=&  \left[ {\rm ln A}\right]^\infty_{r\geq R(t)} 
-4\pi \int_{r\geq R}^\infty A^2 r' \left(_{\rm R}\,\!S^{(r)}_{\,\,\,\,(r)} + 
E_{\rm R} \right)  dr' \,\,\,\,\nonumber \\
&=& {\rm ln }\, A(t,r)^{-1}|_{r\geq R(t)}
-4\pi \int_{r\geq R(t)}^\infty A^2 r' \left(_{\rm R}\,\!S^{(r)}_{\,\,\,\,(r)} + 
E_{\rm R} \right)  dr' \,\,\,\,\nonumber \\
\label{nu*}
&=& {\rm ln} \left[1- \frac{2G_0 m(t,r)}{r}\right]^{1/2}_{r\geq R(t)} 
-4\pi \int_{r\geq R(t)}^\infty A^2 r' \left(_{\rm R}\,\!S^{(r)}_{\,\,\,\,(r)} + 
E_{\rm R} \right)  dr'  \,\,\,\,,
\eea
where we used Eq.(\ref{rgpsA}) in order to replace 
the term with $m/r'^2$ and also the asymptotic flatness condition on $A$ 
[cf. Eq.(\ref{asympt}) below]. 
It is to be stressed that in the absence 
of matter outside the star surface, the first term of Eq.(\ref{nu*})
corresponds to the expression for $\nu$ of the Schwarzshild metric 
(with $m(t,r)|_{r\geq R(t)}= M_*$ being $M_*$ the total mass of the star). 
However, in the present case there are contributions 
(due to the energy-density $E_{\rm R}$  and 
pressure $_{\rm R}\,\!S^{(r)}_{\,\,\,\,(r)}$ of particles which fills 
some part of the space outside the star) which are the responsible 
for the actual metric to deviate from the 
Schwarzschild one. These contributions arise from the second term of 
Eq. (\ref{nu*}). In some cases (e.g. the free streaming 
approximation) numerical analysis show that these deviations are so small 
so that they can be neglected (cf. \cite{gour3}). 

Deviations of this kind can be appreciated more easily in presence 
of non trivial scalar fields, for instance in the phenomenon of 
spontaneous scalarization \cite{ssn}. Moreover, for 
$r> R(t)$, the mass function $m(t,r)$ is larger than $m(t,R(t))$ 
due to the contribution of $E_{\rm R}$ to the total 
energy density [cf. Eq. (\ref{hamrgps2}) ]. Indeed the mass difference 
is given by
\be
\delta m= 4\pi \int_{R(t)}^{r > R(t)} E_{\rm R} r'^2 dr'\,\,\,\,.
\ee

Another way to appreciate such deviations is by noting that Eq. (\ref{AN}) 
together with the asymptotic conditions imply, in the case of vacuum, the 
relationship $AN=1$ which is characteristic of 
the Schwarzschild solution in RGPS coordinates. However, when 
matter is present outside the star, then $AN\neq 1$ (e.g. see \cite{ssn}), 
except of course at spatial infinity.

The boundary conditions for $A$ are similar to those for $N$. Therefore
\be
A(t,0) = A_c(t) \,\,\,\,\\
\,\,\,.
\ee
with $A_c(t)$ such that 
\be\label{asympt}
 A(t,r)_{r\rightarrow \infty} \rightarrow 1\,\,\,.
\ee
 In the case of the 
RGPS coordinates the reparametrization Eq. 
(\ref{adef}) imposes the regularity condition
\be
m(t,0)= 0\,\,\,.
\ee
Since near the origin $m\sim r^3$, the reparametrization inforces 
that $A(t,0)=1$. Then $\partial_t A(t,0)=0$. 
The three metric is thus locally flat at the origin. Moreover, 
provided that the energy-density of sources 
falls off at least as fast as $1/r^4$ outside the star, the same 
mass parametrization will drive $A$ to the required asymptotic value. 
Note, that this behavior of the metric potential $A$ is compatible with the 
regularity and asymptotic conditions for  
$K^r_{\,\,\,\,r}$ that we mentioned above. Moreover, these conditions
 also imply that $J^r$ vanish at the origin as well as asymptotically 
[cf. Eq.(\ref{J_rrgps}) ].

In the case of the IGMS and IGPS coordinates one has an 
elliptic equation 
for $\tilde \alpha$ which is not invariant to a rescaling on $A$. So 
unlike the RGPS coordinates where $A_c=1$, the 
central value $A_c$, should be determined from a shooting method 
or otherwise in order to satisfy the asymptotic flatness condition. 
In fact, near the origin $N^r \sim r$, 
therefore from Eq. (\ref{evalfaIGMS}), it turns that at the origin
\be
\partial_t \alpha(t,0) = {\rm const.}\,\,\,,
\ee
and thus from the definition Eq. (\ref{alfa}) and the 
regularity condition (\ref{regcond}) 
(see below) one concludes that (depending on the sign of the 
constant) $A(t,0)$ can grow exponentially 
\cite{ST80,STP85,ST85b,RST}. The three metric 
is thus conformally flat at the origin in the isotropic gauge. 

In addition to the previous regularity conditions we have also
\be\label{regcond}
\partial_r Q |_{r=0}=0= Q^r|_{r=0} \,\,\,.
\ee
where $Q$ represents collectively the metric potentials and the 
scalar matter field variables 
(e.g, $N,A,m,p,\rho, E$, etc.) and $Q^r$ tensor field components 
(e.g $J^r, K^r_{\,\,\,r}$, etc.).

A convenient way to impose the asymptotic conditions accurately is by 
compactifying the outer domain with the help of 
a transformation $u=1/r$ from the star's surface $r=R$ to spatial 
infinity. In this way the infinite domain $r\in [R,\infty)$ is mapped to 
the compact domain $u\in [1/R, 0)$, so the integration can be 
performed from $1/R$ to ``zero'' with a high degree of accuracy 
(cf. \cite{ssn}). Obviously, the 
field and matter variables for particles are to be matched continously at 
$R$.  

Regarding the distribution function, the regularity condition at the 
center on the particle's radial energy flux is 
$H_{\rm R}=0$ [see Eq. (\ref{Hflux})] which means that the average of the 
particles' radial velocity as measured by the Eulerian observer at the origin 
is zero (local isotropy at $r=0$). Same considerations apply for 
the radial particle-flux-number $j^{(r)}$ [cf. Eq. (\ref{jrRphys})]. 
A sufficient condition for $H_{\rm R}=0=j^{(r)}$ at $r=0$ is 
\be
F_{\rm R}(t,0,e,\mu)= G_{\rm R}(t,e,\mu) \,\,\,{\rm with}\,\,\,G_{\rm R}(t,e,\mu)=
G_{\rm R}(t,e,-\mu)\,\,\,,
\ee
that is, the distribution function being a pair function of $\mu$ at $r=0$, enforces  
that the integrals given by Eqs. (\ref{Hflux}) and (\ref{jrRphys}) 
vanish identically at the origin. This condition ensures in addition
 that with a suitable form of 
$G_{\rm R}(t,e,\mu)$ in the energy domain the energy-density, pressure and 
density number of particles in the Eulerian frame are finite at the star's center. 
Usually the assumption of an ideal quantum gas is adopted as initial 
condition for the particles so that the distribution function is 
isotropic and given by a Fermi-Dirac or Bose-Einstein 
distribution (whether the particles are Fermions or Bosons) \cite{MM,HV,HH}. 
Then the particles energy-density and pressure will be parametrized 
initially only by the thermodynamic variables like the 
temperature. In that case $H_{\rm R}=0$ all over 
the initial spacelike hypersurface. 

The regularity condition for $F_{\rm R}$ at $r=0$ is like other scalar quantities, 
\be
\partial_r F_{\rm R}|_{r=0}=0\,\,\,.
\ee

Another boundary condition is that the inward flux of particles at the star surface is 
zero. This is imposed by \cite{Lind},
\be
F_{\rm R}|_{r=R}= 0 \,\,\,\,{\rm for} \,\,\,-1\leq \mu <0 \,\,\,.
\ee
Returning to the initial conditions, the form of these, will characterize 
first, the type of configuration (e.g., supernova, neutron star, supermassive star, 
star cluster) and second the dynamics and ultimate fate of the precursor 
(e.g.  protoneutron star, neutron star, black hole).  The goal of future numerical 
investigation will be to explore a large set of initial conditions and 
their consequences (cf. \cite{Novak}).

\section{Conclusions}
Several astrophysical phenomena involve the dynamics of relativistic 
objects. Some of the most interesting ones end up in the formation of 
black holes or neutron stars, like the collapse of cores and supernova
explosions. While most of the astrophysical objects are rotating, the 
role of rotation in relativity can be neglected as regards the 
structure of the object when the rotation frequency is low. Aside from 
fast pulsars, most of the astrophysical objects are endowed with a low rotation 
frequency. Therefore, the spherical symmetry is an assumption that can be 
very useful in a wide range of applications. On the other hand, the 
mere existence of fast pulsars reveals that rotation has to be taken 
into acount in a more realistic situation. Moreover, it seems that the deviations from spherical symmetry 
in supernova explosions is central in the phenomenon \cite{BHF}, and that 
rotation can influence the cooling mechanims in early phases of neutron 
stars \cite{Miralles}. In general relativity this is 
a difficult task and only a few attempts have been succeeded within an 
evolutionary code (see \cite{Shibata} and references therein). 

One can separate the problem of the dynamics of relativistic bodies in 
two sets. The first one involves the formulation, the geometry and the numerics. 
A convenient coordinate choice and a powerful numerical analysis can allow long 
term evolutions leading to a better understanding of several physical 
phenomena. Thus this is a crucial point which has been recognized 
by almost all the numerical relativists. Investigations of the 
effect of several gauges and time slicing conditions is always an 
important issue. One of the aims of this paper was therefore to derive 
the fundamental equations under different gauges and write them in 
several forms suitable for different numerical schemes.

The second set, involves the physical approximations used in the model.
In the case of gravitational collapse, we have discussed that neutrinos 
cannot only play an important role in the dynamics but also 
that the signal carried by them can be 
fundamental for a better understanding of the underlying physics and as an 
invaluable imprint of the ultimate fate of the collapsed object. In particular, 
if neutrinos turn to be massive particles \cite{SNO}, mechanisms like the early 
black hole formation could be tested \cite{BBM}.

Furthermore, the equation of state at high densities can also lead to 
different time scale processes during the collapse and the accretion phase.    
Therefore, it turns necessary to pursue the analysis with the incorporation   
of the most recent advances in particle and nuclear physics.

While the formalism presented here included only hydrodynamics and 
transport theory, the equations are quiet general as to include other 
types of matter like scalar fields, 
which are very useful in the analysis of critical phenomena.

Our aim for the future investigations is to perform an extensive numerical 
analysis of several issues discussed here and more generally to analyse the 
dynamics of spherically symmetric spacetimes with several kind of matter 
sources.

\bigskip
\section*{Acknowledgments}
\bigskip

This work has been supported by DGAPA-UNAM, grant No. 112401 
and CONACYT, M\'exico, grant No. 32551-E.


\begin{thebibliography}{}

\bibitem{HWW}
B. K. Harrison, K.S. Thorne, M. Wakano, and J. A. Wheeler, {\it Gravitation 
Theory and Gravitational Collapse}, The University of Chicago Press, 1965

\bibitem{ST83}
S. L. Shapiro, and S.A. Teukolsky, Black Holes, White Dwarfs and Neutron Stars 
(The physics of compact objects), John Wiley \& Sons, New York, 1983

\bibitem{OS}
J. R. Oppenheimer, and H. Snyder, \Journal{\PREV}{56}{455}{1939}

\bibitem{Alcubierre}
M. Alcubierre, G. Allen, B. Br\"ugmann, E. Seidel, and W.M. Suen, 
\Journal{\PRD}{62}{124011}{2000}

\bibitem{Kelly}
B. Kelly {\it et al.}, \Journal{\PRD}{64}{084013}{2001} 

\bibitem{ST80}
S. L. Shapiro, and S.A. Teukolsky, \Journal{\APJ}{235}{199}{1980}

\bibitem{SBP}
P. J. Schinder, S. A. Bludman, and T. Piran, \Journal{\PRD}{37}{2722}{1988}

\bibitem{gour1}
E. Gourgoulhon, \Journal{\AA}{252}{651}{1991}

\bibitem{gour2}
E. Gourgoulhon, \Journal{\APF}{18}{1}{1993}

\bibitem{STP85}
L. I. Petrich, S. L. Shapiro, and S.A. Teukolsky, \Journal{\PRD}{31}{2459}{1985}

\bibitem{STP86}
L. I. Petrich, S. L. Shapiro, and S.A. Teukolsky, \Journal{\PRD}{33}{2100}{1986}

\bibitem{CW}
S. A. Colgate and R. H. White, \Journal{\APJ}{143}{626}{1966}

\bibitem{MW}
M. M. May and R. H. White, \Journal{\JCP}{7}{219}{1967}; {\it ibid} 
\Journal{\PREV}{141}{1232}{1966}

\bibitem{wilson}
J. R. Wilson, \Journal{\APJ}{163}{209}{1971}

\bibitem{MWS}
R. Mayle, J. R. Wilson, and D. N. Schramm, \Journal{\APJ}{318}{288}{1987}

\bibitem{SS}
R. Saenz, and S. Shapiro, \Journal{\APJ}{229}{1107}{1979}

\bibitem{BHF}
A. Burrows, J. Hayes, and B. A. Fryxell, \Journal{\APJ}{450}{830}{1995}

\bibitem{Burrows88}
A. Burrows, \Journal{\APJ}{334}{891}{1988}

\bibitem{BL}
A. Burrows and J. M. Lattimer, \Journal{\APJ}{318}{L63}{1987}

\bibitem{MM}
A. Mezzacappa and R. A. Matzner, \Journal{\APJ}{343}{853}{1989}

\bibitem{Mezzacapa}
A. Mezzacappa, astro-ph/0010579; {\it ibid} astro-ph/0010580

\bibitem{BBM}
J. F. Beacom, R. N. Boyd, and A. Mezzacappa, \Journal{\PRL}{85}{3568}{2000}; 
astro-ph/0010398.

\bibitem{Lieben1}
M. Liebend\"orfer {\it et al.}, \Journal{\PRD}{63}{103004}{2001}

\bibitem{Lieben2}
M. Liebend\"orfer, A.  Mezzacappa, and F. K. Thielemann, 
\Journal{\PRD}{63}{104003}{2001}

\bibitem{MB}
A.  Mezzacappa and S. W. Bruenn, \Journal{\APJ}{405}{669}{1993}; 
{\it ibid} \Journal{\APJ}{405}{637}{1993}, \Journal{\APJ}{410}{740}{1993}

\bibitem{Burrows}
A. Burrows, \Journal{\APJ}{334}{891}{1988}; {\it ibid} 
\Journal{\APJ}{300}{488}{1986}

\bibitem{Baum0}
T. W. Baumgarte, S. L. Shapiro, and S.A. Teukolsky, 
\Journal{\APJ}{443}{717}{1996}

\bibitem{Baum1}
T. W. Baumgarte, S.A. Teukolsky, S. L. Shapiro, H. T. Janka, and W. Keil, 
\Journal{\APJ}{468}{823}{1996}

\bibitem{Janka}
W. Keil and H. T. Janka, \Journal{\AA}{271}{187}{1993}

\bibitem{Baum2}
T. W. Baumgarte, S. L. Shapiro, and S.A. Teukolsky, 
\Journal{\APJ}{458}{680}{1996}

\bibitem{TPL}
V. Thorsson, M. Prakash, and J. M. Lattimer, \Journal{\NPA}{572}{693}{1994}

\bibitem{BB}
G. E. Brown and H. A. Bethe, \Journal{\APJ}{423}{659}{1994}

\bibitem{Glend}
N. K. Glendenning, \Journal{\APJ}{448}{797}{1995}; {\it ibid} 
\Journal{\APJ}{293}{470}{1985}

\bibitem{SNO}
Q.R. Ahmad {\it et al.}, \Journal{\PRL}{87}{071301}{2001}

\bibitem{OMNIS}
D. B. Cline {\it et al.}, \Journal{\PRD}{50}{720}{1994}; 
R. N. Boyd and A. St. J. Murphy, proceedings of NNN99, AIP
Conf Proc 533, (2000) Ed. M Diwan and C.K. Jung, ISBN
1-56396-956-4 

\bibitem{gour3}
E. Gourgoulhon, and P. Haensel, \Journal{\AA}{271}{187}{1993}

\bibitem{ST79}
S. L. Shapiro and S.A. Teukolsky, \Journal{\APJ}{234}{L177}{1979}

\bibitem{ST85b}
S. L. Shapiro and S.A. Teukolsky, \Journal{\APJ}{298}{34}{1985}; {\it ibid} 
\Journal{\APJ}{298}{58}{1985}; {\it ibid} {\it Relativistic Stellar Dynamics On the 
Computer}, in Dynamical Spacetimes and Numerical Relativity, 
(J. M. Centrella ed.), Cambridge University Press, 
England, 1986.

\bibitem{ST85c}
S. L. Shapiro and S.A. Teukolsky, \Journal{\APJ}{292}{L41}{1985}

\bibitem{ST86}
S. L. Shapiro and S.A. Teukolsky, \Journal{\APJ}{307}{575}{1986}

\bibitem{STK}
C. S. Kochanek, S. L. Shapiro, and S.A. Teukolsky, \Journal{\APJ}{320}{73}{1987}

\bibitem{RST}
F. A. Rasio, S. L. Shapiro, and S.A. Teukolsky, \Journal{\APJ}{344}{146}{1989}

\bibitem{Linke}
F. Linke, J.A. Font, H.T. Janka, E. M\"uller, and P. Papadopoulos, 
astro-ph/0103144

\bibitem{HV}
H. Harleston, and E.T. Vishniac, \Journal{\PRD}{45}{4458}{1992}

\bibitem{HH}
H. Harleston, and K. Holcomb, \Journal{\APJ}{372}{225}{1991}  

\bibitem{ADM}
R. Arnowitt, S. Deser, and C.W. Misner, in Gravitation: an introduction to current 
research, (L.Witten ed.), Wiley, New York, 1962

\bibitem{Choquet}
Y. Choquet-Bruhat, Cauchy Problem, in Gravitation: an introduction to 
current research, (L. Witten ed.), Wiley, New York, 1962; 
Y. Choquet-Bruhat, and J. W. York, The Cauchy Problem, in 
General Relativity and gravitation: one hundred years after the birth 
of Albert Einstein, (A. Held ed.), Plenum Press, New York, 1962

\bibitem{msnotes}
M. Salgado, unpublished  notes (see http://www.nuclecu.unam.mx/$\tilde{ }$marcelo)

\bibitem{york79}
J. W. York, {\it Kinematics and dynamics of general relativity}, in 
Sources of Gravitational Radiation, (L. Smarr ed.), Cambridge University Press, 
England, 1979.

\bibitem{Wald}
R. Wald, General Relativity, The University of Chicago Press, Chicago, 1984

\bibitem{Chandra}
S. Chandrasekhar, 1979, in An Einstein Centenary Survey (Hawking S.W, 
W.Israel eds.), Cambridge University Press, Cambridge, 1979; 
S. Chandrasekhar, The Mathematical Theory of Black Holes, Oxford University 
Press, 1983

\bibitem{msrmf}
M. Salgado, \Journal{\RMF}{44}{1}{1998}

\bibitem{msdgfm}
M. Salgado, Proc. of the III Workshop of the DGFM-SMF (eds. N. Bret\'on, 
O. Pimentel, and J. Socorro), Universidad de Guanajuato, 2000.

\bibitem{gour4}
S. Bonazzola and E. Gourgoulhon, in {\it Relativistic Gravitation and Gravitational 
Radiation}, Proc. of the Les Houches School of Physics (eds. J. A. Marck and J. P. 
Lasota), Cambridge University Press, 1997

\bibitem{warn}
In order to avoid any misleading when comparing Eq.(\ref{EulerRG}) with 
Eq. (2.29) of Ref.\cite{gour4}, we point out that 
the term $\,^3\,\!U^{(l)} {\cal O}^{(i)}_{(t)(l)}$ of Eq.(\ref{EulerRG}) 
will arise when transforming the term $n^\mu \nabla_\mu V^\alpha$ of 
Eq. (2.29) of Ref.\cite{gour4} to physical 
components since then $e^{(\sigma)}_\alpha n^\mu \nabla_\mu V^\alpha= 
n^\mu \nabla_\mu V^{(\sigma)} + n^{(\lambda)}V^{(\beta)} 
{\cal O}^{(\sigma)}_{(\lambda)(\beta)}$. Since 
$n^{(t)}=1$, and $V^{(t)}=0= n^{(i)}$, then for $\sigma=i$, we have, 
$n^{(\lambda)}V^{(\beta)} {\cal O}^{(i)}_{(\lambda)(\beta)}= 
V^{(l)}  {\cal O}^{(i)}_{(t)(l)}$.
 
\bibitem{Lind}
R. W. Lindquist, \Journal{\AP}{37}{487}{1966}

\bibitem{Haensel92}
P. Haensel, \Journal{\AA}{262}{131}{1992}

\bibitem{DSI}
R. Duncan, S. Shapiro, and I. Wasserman, \Journal{\APJ}{267}{358}{1983}; 
{\it ibid}, \Journal{\APJ}{278}{806}{1984}

\bibitem{SSS}
E. Salpeter and S. Shapiro, \Journal{\APJ}{251}{311}{1981}; 
P. J. Schinder and S. Shapiro, \Journal{\APJ}{259}{311}{1982}; 
{\it ibid}, \Journal{\APJS}{50}{23}{1982}; 
R. Duncan, S. Shapiro, and I. Wasserman, \Journal{\APJ}{309}{141}{1986}

\bibitem{3+1math}
Most of the algebraic computations which include the 3+1 form of Einstein equations, 
matter equations and the Relativistic Boltzmann equation in spherical symmetry 
were performed by using a MATHEMATICA code presented in 
M. Salgado, \Journal{\CPC}{79}{309}{1994}. 

\bibitem{Schinder}
P. J. Schinder, \Journal{\PRD}{38}{1673}{1988}

\bibitem{SmarYork}
L. Smarr, and J. W. York, \Journal{\PRD}{17}{2529}{1978}

\bibitem{BarPir}
J. M. Bardeen, T. Piran, \Journal{\PR}{96}{205}{1983}

\bibitem{MS}
C. Misner and D. Sharp, \Journal{\PREV}{136}{571}{1964}

\bibitem{Baron}
E. Baron, E. S. Myra, J. Cooperstein, and L. J. Van Den Horn, 
\Journal{\APJ}{339}{978}{1989}

\bibitem{Cern}
J. Cernohorsky and C. G. Van Weert,  \Journal{\APJ}{398}{190}{1992} 
 
\bibitem{Piran}
T. Piran, {\it Methods for Numerical Relativity}, in Rayonnement Gravitationnel 
(ed. N. Deruelle and T. Piran), North Holland, 1983

\bibitem{HM}
W. C. Hernandez and C. W. Misner, \Journal{\APJ}{143}{452}{1966}

\bibitem{Vaidya}
P. C. Vaidya, \Journal{\NAT}{171}{260}{1953}

\bibitem{Lind65}
R.W. Lindquist, R.A. Schwartz, C.W. Misner, \Journal{\PRD}{137}{B1364}{1965}

\bibitem{ssn}
M. Salgado, D. Sudarsky, and U. Nucamendi, \Journal{\PRD}{58}{124003-1}{1998}

\bibitem{Novak}
J. Novak, to be published in Astron. Astrophys. (pre-print gr-qc/0107045)

\bibitem{Miralles}
J. A. Miralles, K. A. Van Ripper, and J. M. Lattimer, 
\Journal{\APJ}{407}{687}{1993} 

\bibitem{Shibata}
M. Shibata, T. W. Baumgarte, and S. L. Shapiro, \Journal{\PRD}{61}{044012-1}{2000}

\end{thebibliography}
\end{document}